\documentclass[aps,prd,review,showpacs,amsmath,amssymb,floatfix,nofootinbib,eqsecnum,notitlepage]{revtex4-1}

\usepackage{graphicx,color,framed}
\usepackage{hyperref}
\usepackage{times}
\usepackage{enumerate}
\usepackage{lipsum}
\usepackage{slashed}
\usepackage{array}
\usepackage{textcomp}
\usepackage{amsthm}
\usepackage{comment}

\hypersetup{
    colorlinks=true, 
    linktoc=all,     
    linkcolor=blue,  
}

\newtheorem{thm}{Theorem}
\newtheorem*{thm*}{Theorem}

\newtheorem{lemma}{Lemma}
\newtheorem*{lemma*}{Lemma}
\newtheorem{claim}{Claim}

\def \beq {\begin{equation}}
\def \eeq {\end{equation}}
\def \beqa {\begin{eqnarray}}
\def \eeqa {\end{eqnarray}}
\def \bseq {\begin{subequations}}
\def \eseq {\end{subequations}}
\newcommand \dg {\dagger}
\newcommand \up {\uparrow}
\newcommand \down {\downarrow}
\newcommand \ran {\rangle}
\newcommand \lan {\langle}

\newcommand \ep {\epsilon}

\newcommand \mb {\mathbf}

\newcommand \nnb {\nonumber}

\newcommand \ov {\overline}
\newcommand \td {\tilde}

\newcommand \al {\alpha}
\newcommand{\bpm}{\begin{pmatrix}}
\newcommand{\epm}{\end{pmatrix}}
\newcommand{\bmm}{\begin{matrix}}
\newcommand{\emm}{\end{matrix}}

\begin{document}

\title{Stability of ground state degeneracy to long-range interactions}

\author{Matthew F. Lapa}
\email[email address: ]{matthew.lapa@mailaps.org}
\affiliation{Kadanoff Center for Theoretical Physics, University of Chicago, Chicago, IL, 60637, USA}

\author{Michael Levin}
\email[email address: ]{malevin@uchicago.edu}
\affiliation{Kadanoff Center for Theoretical Physics, University of Chicago, Chicago, IL, 60637, USA}


\begin{abstract}

We show that some gapped quantum many-body systems have a ground state degeneracy that is stable to long-range (e.g., power-law) perturbations, in the sense that any ground state energy splitting induced by such perturbations is exponentially small in the system size.
More specifically, we consider an Ising symmetry-breaking Hamiltonian with several exactly degenerate ground states and an energy gap, and we then perturb the system with Ising symmetric long-range interactions. For these models we prove (1) the stability of the gap, and (2) that the residual splitting of the low-energy states
below the gap is exponentially small in the system size. Our proof relies on a convergent polymer expansion that is adapted to handle the long-range interactions in our model. We also discuss applications of our result to several models of physical interest, including the Kitaev p-wave wire model perturbed by power-law density-density interactions with an exponent greater than 1. 

\end{abstract}


\maketitle

\section{Introduction}

Some gapped quantum many-body systems have the interesting property that they have multiple nearly degenerate ground states with an energy splitting that is \emph{exponentially} small in the system 
size~\cite{wen-niu,kitaev-toric-code,kitaev}. Furthermore, this nearly exact degeneracy is a \emph{robust} phenomenon in the sense that it does not require fine tuning parameters in the Hamiltonian. 
There is an ongoing effort to realize systems of this kind in the laboratory, as many of them have been argued to be useful platforms for a reliable quantum memory or even a fault-tolerant quantum 
computer~\cite{TQC}. 

This kind of nearly exact ground state degeneracy is a well-established phenomenon for systems with \emph{short-range} interactions (e.g., finite-range interactions). In this case, there are rigorous arguments proving the existence of an exponentially small ground state splitting in many concrete models~\cite{kirkwood-thomas,klich2010,BHM,BH,michalakis-zwolak,nachtergaele2019,nachtergaele2020}. On the other hand, much less is known about exponentially small ground state splitting for systems with long-range interactions (e.g., power-law decaying interactions). This is despite the fact that long-range interactions are present in many candidate systems with ground state degeneracy, such as topological superconductors and fractional quantum Hall liquids~\cite{kitaev,wen-niu}. The main purpose of this paper is to present a rigorous result showing that a robust, exponentially small ground state splitting also occurs
in some models with long-range interactions.

The simplest place to study these issues is in the context of perturbations of exactly solvable models. Suppose $H_0$ is an exactly solvable short-range Hamiltonian, and suppose further that $H_0$ has several \emph{exactly} degenerate ground states which are separated from the rest of the spectrum by a finite energy gap. Consider a generic perturbation of $H_0$ of the form
\beq
	H = H_0 + \lambda V \nnb \ ,
\eeq
where $\lambda$ is a real coefficient and $V$ is an interaction with both a short-range and a long-range part. In this context, our question becomes one of \emph{stability}: in particular, does the energy gap stay open when we turn on a small $\lambda$, and if so, how large is the residual splitting of the states below the gap?  

For short-range perturbations, an exponential bound on this residual splitting has been obtained
in particular cases in Refs.~\onlinecite{kirkwood-thomas,klich2010}, and in great generality for
topologically-ordered $H_0$'s in 
Refs.~\onlinecite{BHM,BH,michalakis-zwolak,nachtergaele2019,nachtergaele2020}.\footnote{In fact, in 
Refs.~\onlinecite{BHM,BH,michalakis-zwolak,nachtergaele2019,nachtergaele2020} the decay is slightly slower 
than a pure exponential, but still faster than any inverse power of the system size.} 
These papers are all part of a much larger collection of 
works~\cite{KT,borgs1996,datta1996,froehlich-pizzo,kirkwood-thomas,datta-kennedy,bravyi-SW,
yarotsky2006,klich2010,BHM,BH,michalakis-zwolak,nachtergaele2019,nachtergaele2020,
hastings-fermi,de-roeck-salmhofer,koma-fermions} that focused
on proving various stability results for gapped Hamiltonians subjected to short-range perturbations 
(Ref.~\onlinecite{michalakis-zwolak} also considered long-range perturbations -- see the next paragraph).
The papers in this collection utilize a variety of different methods and apply to a variety of different
kinds of unperturbed Hamiltonians, for example the case where $H_0$ is a classical 
Hamiltonian~\cite{KT,borgs1996,datta1996} and
the case where $H_0$ is a quantum Hamiltonian obeying a \emph{Topological Quantum Order} 
condition~\cite{BHM,BH,michalakis-zwolak,nachtergaele2019,nachtergaele2020}.

In contrast, when $V$ is a long-range perturbation, say containing power-law interactions, it is not even clear that one should expect an exponentially small bound on the residual splitting of the low-energy
states. For example, one might guess that power-law interactions would generically lead to a power-law bound on the residual splitting (i.e., the splitting is bounded from above by a constant times the system size raised to a negative power). Indeed, the only known stability results in the power-law case were obtained in Ref.~\onlinecite{michalakis-zwolak}, and those results imply a power-law bound on the residual splitting.

To gain some intuition on the difficulty of obtaining an exponential splitting bound in the long-range case, it is useful to review two of the main ways of arguing for the exponential splitting bound in the case of a short-range perturbation. For simplicity of exposition, we review these methods in the case of the one dimensional transverse field Ising chain of length
$L$, where $H_0 = -\sum_{j=1}^L \sigma^z_j\sigma^z_{j+1}$ and $V = \sum_{j=1}^L \sigma^x_j$. In this case $H_0$ has
two exactly degenerate low-energy states given by $|\pm\ran = (|\Uparrow\ran \pm |\Downarrow\ran)/\sqrt{2}$,
where $|\Uparrow\ran$ and $|\Downarrow\ran$ are the states of the chain with all spins up and all spins down,
respectively. 

The first method relies on the technique of \emph{quasiadiabatic continuation} (QAC) and was already discussed in the original paper on that technique~\cite{QAC}. This method can be used to prove an exponential splitting bound when the perturbation parameter $\lambda$ is small enough so that the energy gap of the system stays open. The basic idea is as follows: given our assumption that the energy gap of $H$ stays open, $H$ will have two low-energy states $|\pm,\lambda\ran$ that evolve out of the unperturbed states $|\pm\ran$. Quasiadiabatic continuation\footnote{Actually, what we are describing here is known as \emph{exact} QAC and was developed starting in Ref.~\onlinecite{Osborne}. For an introduction to exact QAC, see Ref.~\onlinecite{hastings-locality}.} allows one to construct a unitary operator $U_{\lambda}$ with the following two properties:
\begin{enumerate}
\item $U_{\lambda}$ maps the low-energy states of $H_0$ to those of $H$: $|\pm,\lambda\ran = U_{\lambda}|\pm\ran$.
\item Conjugation by $U_{\lambda}$ takes local operators to local operators, up to a rapidly decaying
tail (the decay is faster than any power in the distance).
\end{enumerate}
One can then express the energy splitting $\delta$ between the states $|+,\lambda\ran$ and
$|-,\lambda\ran$ as
\beqa
	\delta &=& \lan +|U_{\lambda}^{\dg}HU_{\lambda}|+\ran - \lan -|U_{\lambda}^{\dg}H U_{\lambda}|-\ran \nnb \\
	&=& 2\text{Re}\left\{ \lan \Uparrow|U_{\lambda}^{\dg}H U_{\lambda}|\Downarrow\ran   \right\}\ .
\eeqa
We see that the energy splitting $\delta$ evaluates to the matrix element of the transformed Hamiltonian 
$U_{\lambda}^{\dg}H U_{\lambda}$ between the all-up and all-down states. This expression is quite
intuitive as it is reminiscent of an instanton-type tunneling effect. Note that, for any operator
$\mathcal{O}$, $\lan\Uparrow|\mathcal{O}|\Downarrow\ran = 0$ unless
$\mathcal{O}$ acts nontrivially on all $L$ sites of the chain. Since $H$ is a sum of
local terms, the part of  $U_{\lambda}^{\dg}H U_{\lambda}$ that acts on all $L$ sites
decays rapidly with $L$ 
by the properties of the QAC operator $U_{\lambda}$. Therefore we find that $\delta$ decays faster than any power of $L$,
which is very close to the exponential splitting result that we wanted to prove.\footnote{It is possible
to prove a true exponential splitting bound using the approximate (or Gaussian) version of QAC
as in the original paper~\cite{QAC}.}

While the QAC method works very well for short-range interactions, it has serious limitations for long-range interactions such as power-law interactions. The problem is that conjugation by $U_\lambda$ takes local operators to local operators with power-law (or larger) tails. Therefore, the QAC-based method only seems capable of proving a \emph{power-law} splitting bound in the presence of power-law interactions. Indeed, the results of Ref.~\onlinecite{michalakis-zwolak} are based on this technique and, as we mentioned above, in the power-law case they imply a power-law bound on the residual splitting. 

The other well-known method for arguing for an exponential splitting is an intuitive argument based on perturbation theory~\cite{kitaev-toric-code}. The basic idea is that the matrix elements $\lan \Uparrow |V^p|\Downarrow\ran$ of the $p$th power of $V$ between the all-up and all-down states vanish for all powers $p$ until $p = L$, the length
of the chain. Hence, the energies of the states $|+\ran$ and $|-\ran$ will not deviate until order 
$L$ in perturbation theory in $\lambda$, which suggests that the splitting between these two states will scale as $\lambda^L$ and will therefore be exponentially small in the system size. This argument can be made precise if one can show that the perturbative expansion for the ground state energy has a finite radius of convergence in the thermodynamic limit. One way to prove results of this nature is to use a convergent ``polymer expansion'', a type of perturbation theory for the free energy of a model. This method was used in the stability proofs in Refs.~\onlinecite{KT,borgs1996,datta1996,yarotsky2006,klich2010} for systems with short-range interactions. 

Like the QAC method, the standard perturbative method encounters difficulties for systems with 
long-range interactions. 
The problem is that the combinatorial arguments
that guarantee the convergence of the polymer expansion typically rely on bounding the number of polymers of a particular size that contain a given space-time point, and the presence of long-range interactions invalidates these bounds by vastly increasing the number of possible polymers.\footnote{The standard combinatorial arguments for systems with short-range 
interactions are reviewed in Ch.~V of Ref.~\onlinecite{simon}.} Fortunately, this difficulty can be overcome with more sophisticated combinatorial arguments. Indeed, in the closely related problem of \emph{classical} statistical mechanical models with long range interactions, convergent polymer expansions have been established in a number of systems~\cite{ginibre1966condensation,kunz1978analyticity,israel-nappi,
frohlich-spencer,
imbrie,cammarota1982decay,park1988extension,cassandro2005geometry,procacci2007abstract,affonso2021long}.

In this paper, we put these ideas together and extend the perturbative method to a class of quantum systems with long-range interactions.
We focus on a simple case where $H_0$ is a classical symmetry-breaking Hamiltonian with several 
exactly degenerate low-energy states that are separated from the rest of the spectrum by a finite energy gap.
Thus, our initial setting is similar to Refs.~\onlinecite{KT,borgs1996,datta1996,froehlich-pizzo,
kirkwood-thomas,datta-kennedy}.
We then add to this initial Hamiltonian a (symmetric) perturbation $\lambda V$ where $V$ consists of both 
short- and long-range parts,
\beq
	V = V_{\text{short}} + V_{\text{long}}\nnb \ .
\eeq 
In the long-range part of $V$, we allow a large class of power-law and other long-range interactions, including not only $2$-body terms but also more general $K$-body terms (for $K$ of order $1$).
For these Hamiltonians, and
for $\lambda$ small enough, we prove (1) stability of the spectral gap, and (2) an exponentially 
small (in system size) bound on the residual splitting of the low-energy states below the gap. Thus, we are
able to establish an exponential splitting bound for a large family of Hamiltonians with long-range 
interactions. We also apply our results to several models of current interest 
including Kitaev's p-wave wire model~\cite{kitaev} perturbed by 
power-law density-density interactions, as well as a number-conserving model of a topological superconducting 
qubit~\cite{LL,LL2}.

To conclude this introduction, we highlight several previous works that studied
gapped and/or topological phases in the presence of long-range interactions. First, several works have
presented evidence for the existence of gapped topological phases in the presence of long-range (usually
power-law) interactions~\cite{PhysRevB.83.155110,PhysRevB.87.081106,PhysRevB.90.085146}. 
In addition, there is a large body of work, beginning with
Ref.~\onlinecite{PhysRevB.88.155420}, that focused on the effects of long-range hopping and pairing terms 
in free fermionic systems (in one and higher dimensions) as well as related Ising spin chains (in the
one-dimensional case). For our purposes here, 
Refs.~\onlinecite{PhysRevLett.113.156402,vodola2015long,PhysRevLett.118.267002,
PhysRevB.93.041102,PhysRevB.93.205115,PhysRevA.91.032303} are of particular interest, as all of these
works presented negative results for exponentially small 
ground state degeneracy splitting in systems with long-range interactions, including some
systems where it was possible to check the stability of the spectral gap (i.e., stability of the phase).
In contrast to these works, we have been able to establish 
not only the stability of the spectral gap in our models, but also the exponentially 
small bound on the residual splitting of the low-energy states below the gap.

This paper is organized as follows. In Sec.~\ref{sec:model-and-main-result} we introduce 
an example of the kind of model that our result applies to, and then we state our main result in
the context of this model. In Sec.~\ref{sec:main-proof} we prove our main result for this model and 
discuss the physical intuition that underlies the formal proof. In Sec.~\ref{sec:generalizations} we
explain how our main result can be generalized to a much larger family of models than we considered
in Secs.~\ref{sec:model-and-main-result} and ~\ref{sec:main-proof}. Section~\ref{sec:conclusion} presents our
conclusions. Finally Appendixes~\ref{app:weights}, \ref{app:combinatorics}, \ref{app:polymer} 
and \ref{app:generalizations} contain important details for the proofs of
our main results from Sections \ref{sec:main-proof} and \ref{sec:generalizations}.

\section{Stability of the transverse field Ising model to long-range interactions}
\label{sec:model-and-main-result}

In this section we state our main result in the context of a prototypical model that is both
complicated enough to contain all of the physics we are interested in, and simple enough
to allow for a straightforward demonstration of our method of proof. In Sec.~\ref{sec:generalizations} we 
explain how our result can be generalized to a much larger family of models.

\subsection{Description of the model and its symmetry}
\label{sec:model}

We consider a one dimensional spin-$1/2$ chain with $L$ sites and periodic boundary conditions, and we assume that 
$L$ is even for convenience, although our results would also hold without this assumption. The
sites on the chain are labeled by $j\in\{1,\dots,L\}$, and on each site
we have the three Pauli operators $\sigma^x_j$, $\sigma^y_j$, and $\sigma^z_j$. We measure distances
on the chain using the periodic distance function, $|j-k|_{\text{p}} := \min_{n\in\mathbb{Z}}|j-k + nL|$. The Hamiltonian for our model takes the form
\beq
	H= H_0 + \lambda V\ ,
\eeq
where $H_0$ is a classical Ising Hamiltonian,
\beq
	H_0 = \frac{\Delta}{2}\sum_{j=1}^{L}(1-\sigma^z_j\sigma^z_{j+1})\ ,
\eeq
and where $\Delta>0$ is an overall energy scale and 
$\sigma^{x,y,z}_{L+1}\equiv \sigma^{x,y,z}_1$ because of the periodic boundary conditions. The perturbation term $V$ takes the form
\beq
	V= h\sum_{j=1}^L  \sigma^x_j + \frac{1}{2}\sum_{j\neq k}f(|j-k|_{\text{p}})\sigma^x_j \sigma^x_k\ ,
\eeq
where $h$ is a dimensionless real parameter and
$f(r)$ is a dimensionless real function that determines the long-range interaction. Note that the first term 
here (with coefficient $h$) is the usual transverse field term.

As we mentioned above, the function $f(r)$ determines the long-range part of $V$.
Throughout the paper we assume that $f(r)$ satisfies two conditions. First, we assume that $f(r)$ is
normalized so that $|f(r)| \leq 1$ for all $r$.
Next, we assume that $f(r)$ satisfies a \emph{summability condition} of the form
\beq
	\sum_{r=1}^{\frac{L}{2}}|f(r)| \leq \frac{c}{2}\ , \label{eq:sum-bound}
\eeq
with a constant $c$ that can be chosen to be independent of $L$.\footnote{The upper bound of 
$L/2$ on the range of this sum is due to the fact that $|j-k|_{\text{p}}\leq L/2$ for all
$j$ and $k$.} A typical example of a long-range interaction that satisfies this condition is 
a power law interaction $f(r) = \frac{1}{r^{\al}}$, provided that the power $\al$ satisfies $\al > 1$ (note the strict inequality).

Our assumption that $c$ can be chosen to be independent of $L$ is crucial for
our main result. This assumption guarantees that the long-range term in the Hamiltonian has an
operator norm that is \emph{extensive} (i.e. linear in $L$):  
\beqa
	\Big|\Big| \frac{1}{2}\sum_{j\neq k}f(|j-k|_{\text{p}})\sigma^x_j \sigma^x_k \Big|\Big| 
&\leq& \frac{c L}{2}\ . 
\eeqa

Later in this section, we demonstrate the importance of the summability condition \eqref{eq:sum-bound}, by giving an example of an \emph{instability} for a long-range interaction that violates this condition.

Just like the usual transverse field Ising model, our model has a $\mathbb{Z}_2$ Ising symmetry, generated by the operator
\beq
	\mathcal{S}=\prod_{j=1}^L\sigma^x_{j}\ . \label{eq:ising-sym}
\eeq
The Hilbert space $\mathcal{H}$ of our spin chain can be
divided into two sectors, which we denote by $\mathcal{H}_{+}$ and $\mathcal{H}_{-}$, such
that any state $|\psi_{\pm}\ran\in\mathcal{H}_{\pm}$ satisfies 
$\mathcal{S}|\psi_{\pm}\ran=\pm|\psi_{\pm}\ran$. Then, since $H$ commutes with $\mathcal{S}$, we
can separately diagonalize $H$ within each sector.

\subsection{Main result}

To state our main result, we need to review three key properties of the unperturbed Hamiltonian $H_0$. The first property of $H_0$ is that it has a unique ground state in each sector $\mathcal{H}_\pm$. Specifically, the ground state in the sector $\mathcal{H}_\pm$ is the state $|\pm\ran = \frac{1}{\sqrt{2}}(|\Uparrow\ran \pm |\Downarrow\ran)$ where $|\Uparrow\ran$ denotes the state with all spins pointing up ($\sigma^z_j=1\ \forall\ j$), and $|\Downarrow\ran$ denotes the state with all spins pointing down ($\sigma^z_j=-1\ \forall\ j$). The second property of $H_0$ is that it has a finite energy gap ($=2\Delta$) above the ground state within each sector. The final property of $H_0$ is that the two ground states $|+\ran$ and $|-\ran$ are exactly degenerate. 

Our main stability result says that $H$ possesses approximately these same properties, in the limit $L \rightarrow \infty$, for small but finite values of $\lambda$:

\begin{thm}
\label{thm:stability}
There exists a $L$-independent constant $\lambda_0>0$
such that, if $|\lambda| < \lambda_0$, then (1) $H$
has a unique ground state and a finite energy gap in each sector $\mathcal{H}_{\pm}$ of the Hilbert
space with fixed $\mathcal{S}$ eigenvalue, and (2) the ground
state energy splitting $|E_{+}(\lambda)-E_{-}(\lambda)|$ between sectors satisfies the exponential bound
\beq
	|E_{+}(\lambda)-E_{-}(\lambda)| \leq c_1 L e^{-c_2 L}\ , \label{eq:exponential-bound}
\eeq
where $c_1$ and $c_2$ are positive constants that depend on $\Delta$, $h$, and $\lambda$, 
but not on $L$. 
\end{thm}

\subsection{Applications to other models}

We now discuss some applications of Theorem~\ref{thm:stability} to important models in condensed matter physics. The first model that we discuss is Kitaev's p-wave wire (KpW)
model~\cite{kitaev}.  
The degrees of freedom in this model are spinless fermions hopping on 
the sites $j\in\{1,\dots,L\}$ of a one-dimensional chain. The Hamiltonian with open boundary conditions takes 
the form
\begin{align}
	H_{\text{KpW}} =& -\sum_{j=1}^{L-1}\left( \frac{t}{2} c^{\dg}_{j} c_{j+1} + \frac{\Delta}{2}c_{j} c_{j+1} +\text{h.c.}\right) - \mu \sum_{j=1}^L\left(n_{j} -\frac{1}{2}\right)\ ,
\end{align}
where $c_j$ and $c^{\dg}_j$ are the spinless fermion operators, $n_j = c^{\dg}_j c_j$, and $t,\Delta,$ and
$\mu$ are the hopping energy, pairing energy, and chemical potential, respectively.

This model is in its topological phase for $|\mu| < t$ and $\Delta\neq 0$. The
special point $t = \Delta$ and $\mu = 0$ (where the Majorana zero modes at the ends decouple from
the bulk) is known to have the following stability properties when perturbed 
by a generic weak short-range interaction $V_{\text{short}}$: \footnote{This can be proven using the general
results from Ref.~\onlinecite{BHM}. Reference~\onlinecite{bravyi-koenig} contains an 
alternative proof that holds for the case of quadratic perturbations only.} 
\begin{enumerate}
\item $H = H_{\text{KpW}} + V_{\text{short}}$ has a unique ground state and a finite energy gap in 
each sector of the fermion Fock space with fixed fermion parity.
\item Let $E_{\pm}$ be the ground state energy of $H$ in the sector with fermion parity equal to $\pm 1$. Then, 
$|E_{+}-E_{-}|\leq c_1 L e^{-c_2 L}$, for some $L$-independent constants $c_1$ and $c_2$.
\end{enumerate}

A major open question about this model is whether these stability properties persist in the
presence of long-range interactions. As we mentioned in the introduction, 
Refs.~\onlinecite{PhysRevLett.113.156402,vodola2015long,PhysRevLett.118.267002} have provided evidence
that the exponential bound on the ground state energy splitting does not survive in the presence of 
long-range hopping and pairing terms. Our result
does not apply to those models, but it does apply to the arguably more physical case of a
long-range density-density interaction of the form
\beq
	V_{\text{long}} = \frac{\lambda}{2}\sum_{j\neq k}f_{jk}n_j n_k\ .
\eeq 
Here $\lambda$ is a coupling constant and $f_{jk}= f_{kj}$ is a two-body interaction whose absolute value 
is bounded from above by a positive function $g(|j-k|)$ of the distance between two sites, 
$|f_{jk}|\leq g(|j-k|)$, with $g(r)$ satisfying the summability condition $\sum_r g(r) \leq c$ for some constant $c$.

To see why our results apply to this model, note that at the special point where $t=\Delta$, the perturbed KpW Hamiltonian 
$H = H_{\text{KpW}} + V_{\text{long}}$ can be mapped, via a Jordan-Wigner transformation, 
onto a spin model of the form
\beq
	H = -\frac{\Delta}{2}\sum_{j=1}^{L_1}\sigma^z_j\sigma^z_{j+1} + \sum_{j=1}^L h_j\sigma^x_j + \frac{\lambda}{8}\sum_{j\neq k} f_{jk}\sigma^x_j \sigma^x_k + \text{constant}\ ,
\eeq
where the transverse fields $h_j$ are given in terms of  the parameters $\mu$ and 
$\lambda$ by $h_j = \frac{\mu}{2} - \frac{\lambda}{2}\sum_{k\neq j}f_{jk}$. 
This model is clearly very similar to the spin model that we study in this paper, up to a breaking of 
translation invariance and a redefinition of the parameters. 

Our stability result does indeed apply to this model, after some minor modifications to accommodate the 
breaking of translation invariance. Therefore, we conclude that the energy gap and exponentially 
small ground state splitting of the KpW model survives in the presence of sufficiently weak 
long-range density-density interactions $V_{\text{long}}$. We can also extend this result to the case where $t \neq \Delta$, by thinking of the deviation from the point $t = \Delta$ as an additional (short-range) perturbation. For more details on this we refer the reader to our discussion on generalizations of our result in Sec.~\ref{sec:generalizations}.

Our second example relates to the subject of \emph{number-conserving} models
of topological superconductivity. We will explore this example in more detail in a separate paper~\cite{LL2},
and so we only give a brief description here. Most studies of topological superconductivity rely on a mean-field description of superconductivity -- a description that violates the physical symmetry
of particle number conservation. Recently, several authors have argued that this breaking of 
number conservation could pose a problem for the proposed applications of topological superconductors
to quantum computation. In Ref.~\onlinecite{LL} we began an investigation of this issue
in the context of a number-conserving version of the KpW model that features an additional degree of freedom
that represents a bulk s-wave superconductor. This additional degree of freedom can exchange Cooper pairs
with the fermionic wire, and the total number of fermions in the model is conserved (see
Ref.~\onlinecite{LL} for the details of the model). 
In Ref.~\onlinecite{LL} we proved one stability property of this model, namely the 
existence of a unique ground state and a finite energy gap in each sector of the Hilbert space with fixed 
\emph{total particle number}.

One issue that was not addressed in Ref.~\onlinecite{LL} was the size of the ground state
degeneracy splitting in a two-wire ``qubit'' setup in which the model has \emph{two} low-energy states
below a finite energy gap in each sector of fixed total particle number. Using our main
result in this paper, we can now address this question: our results imply that the residual splitting between
these two low-energy states is exponentially small in the length $L$ of the wires. The key to applying
our result to that model is a mapping, which holds in each sector of fixed total particle number, that
maps our number-conserving topological superconductor model to a spin model with an all-to-all 
long-range interaction with $f(r) = 1/L$ for all separations $r$ (this interaction originates
from the charging energy term in the model from Ref.~\onlinecite{LL}).

\subsection{Instability for interactions that violate the summability condition}

To close this section, we now present a simple variational argument that shows why the summability condition (\ref{eq:sum-bound}) is necessary for our stability result. This argument establishes an \emph{instability} for models of the same form as our Hamiltonian but where $f(r)$ no longer satisfies \eqref{eq:sum-bound}. The instability that we 
discuss appears for \emph{negative} values of $\lambda$ of arbitrarily small magnitude. As in a famous
argument of Dyson~\cite{dyson}, this instability is strong evidence that any perturbation theory 
in $\lambda$ has a vanishing radius of convergence when $f(r)$ does not satisfy the summability 
condition. 

To demonstrate the instability, we compute the expectation value of $H$ in two different variational 
trial states. For the first trial state we choose the ``ferromagnetic'' state $|\Uparrow\ran$ that has all of the 
spins pointing in the positive $z$-direction: $\sigma^z_j|\Uparrow\ran = |\Uparrow\ran$ for all $j$. 
For the second trial state we choose the ``paramagnetic'' state $|\Rightarrow\ran$ that  has all of the 
spins pointing in the positive $x$-direction: $\sigma^x_j|\Rightarrow\ran = |\Rightarrow\ran$ for all $j$. 
If the variational calculation shows that $|\Rightarrow\ran$ is favored over $|\Uparrow\ran$, 
even for very small values of $|\lambda|$, then we will
interpret that finding as indicating an \emph{instability} of $H_0$ to the interaction $V$. 

Following this plan, and setting $h=0$ for simplicity, we find
\begin{align}
\lan\Uparrow|H|\Uparrow\ran = 0 \ , \quad \quad \quad
	\lan\Rightarrow|H|\Rightarrow\ran = \frac{\Delta}{2}L + \frac{\lambda}{2} \sum_{j \neq k} f(|j-k|_{\text{p}})\ .
\end{align}
To proceed, let us consider the case where the interaction $f(r)$ is \emph{positive} and does not satisfy Eq.~\eqref{eq:sum-bound}. For example, consider the standard Coulomb interaction, $f(r) = 1/r$. With 
this choice we find that
\beq
	\sum_{j \neq k} f(|j-k|_{\text{p}}) \sim c' L\ln(L)\ 
\eeq
for some positive constant $c'$. Taking $\lambda$ to be \emph{negative}, the energy of the trial state 
$|\Rightarrow\ran$ is then
\beq
	\lan\Rightarrow|H|\Rightarrow\ran \sim \frac{\Delta}{2}L - \frac{c'|\lambda|}{2}L\ln(L)\ .
\eeq

Because of the additional $\ln(L)$ factor, the paramagnetic trial state $|\Rightarrow\ran$ is
\emph{always} (for $\lambda <0$) a better trial state than the ferromagnetic state $|\Uparrow\ran$ for large 
enough system size $L$. On the other hand, for an interaction that does satisfy the summability 
condition we instead find that 
\beq
	\lan\Rightarrow|H|\Rightarrow\ran \geq \frac{\Delta}{2}L - \frac{c|\lambda|}{2}L\ ,
\eeq
and so in this case the ferromagnetic state $|\Uparrow\ran$ is always a better choice for 
$|\lambda| < \Delta/c$.
These results clearly show that the summability condition \eqref{eq:sum-bound} is necessary for a general stability
result like Theorem~\ref{thm:stability}.

\section{Proof of the main result}
\label{sec:main-proof}

In this section we present the proof of Theorem~\ref{thm:stability}. We begin with an outline of the proof, summarizing the two main steps. We then explain these steps in detail in Secs.~\ref{sec:step1} - \ref{sec:step2}.

\subsection{Outline of the proof}

Our proof of Theorem~\ref{thm:stability} can be divided into two steps.
In the first step, we study the partition function of our model at inverse temperature
$\beta$ in each sector $\mathcal{H}_{\pm}$ of the Hilbert space. We define partition functions $Z_{\pm}$ by
\beq
	Z_{\pm} = \text{Tr}_{\pm}\{e^{-\beta H}\}\ ,
\eeq
where $\text{Tr}_{\pm}(\cdots)$ denotes a trace over $\mathcal{H}_{\pm}$. For later use we also define
the unperturbed partition function $Z_{0}$ in each sector by
\beq
	Z_{0} = \text{Tr}_{+}\{e^{-\beta H_0}\} = \text{Tr}_{-}\{e^{-\beta H_0}\}\ .
\eeq

In the context of quantum statistical mechanics, one can often express partition functions as a sum over configurations (usually worldlines of some kind) 
on an appropriate spacetime region. In the first step of the proof we develop a precise representation of
this kind by showing that $Z_{\pm}$ can be written in the form
\beq
	\frac{Z_{\pm}}{Z_0} = \sum_{X}W_{\pm}(X)\ , \label{eq:Z-as-sum}
\eeq
where the sum is taken over a \emph{finite} set of geometric configurations $X$ on the spacetime region 
$[1,L+1)\times[0,\beta)$, and $W_{\pm}(X)\in\mathbb{C}$ is a complex
weight for the configuration $X$. In what follows we refer to the configurations $X$ as 
\emph{support sets}. This part of our proof closely follows Kennedy and Tasaki (KT)~\cite{KT}.

The support sets $X$ are constructed from a set of basic building blocks that
includes ``boxes'', ``plaquettes'', and ``dashed lines.'' In addition, the
support sets $X$ that appear in the expression for $Z_{\pm}$ all have the further property that 
they can be decomposed into a union of  
non-intersecting\footnote{We give a precise definition of the notion of intersection for these 
support sets later in this section.} elementary support sets $\chi_1,\chi_2,\dots$,
\beq
	X = \chi_1 \cup \chi_2 \cup \cdots  \label{eq:config-factors}
\eeq 
where the weights $W_{\pm}(X)$ factor as
\beq
	W_{\pm}(X) = W_{\pm}(\chi_1)W_{\pm}(\chi_2)\cdots \label{eq:weight-factors}
\eeq
For reasons that we explain later, we refer to $\chi_1,\chi_2,\dots$ as \emph{weakly-connected} support sets.

Note that the weights $W_\pm(X)$ will generally depend on $\lambda$ (as well as the other parameters in $H$).
In what follows, we will allow this parameter $\lambda$ to take on \emph{complex} values: $\lambda \in\ \mathbb{C}$. 
Thus, we are now studying a ``complexified'' 
version of $Z_{\pm}$ whose restriction to real values of $\lambda$ is equal to the partition function for our original model.

Next, in order to gain analytic control over the expansion (\ref{eq:Z-as-sum}), we prove an upper bound on the
absolute values of the weights $W_{\pm}(X)$. 

\begin{lemma}
\label{lem:weights}
For any $\mu,\delta >0$ there exists a $L$- and $\beta$-independent constant $\lambda_0>0$
such that, if $|\lambda|\leq \lambda_0$, then the weight 
$W_{\pm}(X)$ for any support set $X$ (not necessarily weakly-connected) satisfies the bound
\beq
	|W_{\pm}(X)| \leq e^{-\mu |X|}\delta^{d(X)}\left[\prod_{r=1}^{\frac{L}{2}}|f(r)|^{d_r(X)} \right] \ ,
	\label{eq:weight-bound}
\eeq
where $|X|$ is the number of boxes and plaquettes in $X$, $d_r(X)$ is the number of dashed lines of length 
$r$ in $X$, and $d(X)=\sum_{r=1}^{\frac{L}{2}}d_r(X)$.
\end{lemma}

To summarize, the key results from the first step of the proof are Eqs.~\eqref{eq:Z-as-sum}, 
\eqref{eq:config-factors}, and \eqref{eq:weight-factors}, and Lemma~\ref{lem:weights}.

In the second step we use the representation of $Z_{\pm}/Z_0$ from 
\eqref{eq:Z-as-sum} to develop
an expansion, known as a \emph{polymer expansion}, for the logarithm, $\ln(Z_{\pm}/Z_0)$. The
main difficulty in this step is to prove that our expansion for $\ln(Z_{\pm}/Z_0)$ is absolutely 
convergent (for small enough $|\lambda|$). This is the most important part of our proof and 
it is the part where the long-range nature of the interactions in our model really comes into 
play. As we review below, the key to proving the convergence of the polymer expansion is the following lemma, 
which we prove as part of the second step of our proof. To state the lemma, we first define a modified weight $\td{W}(X)$ using the upper bound in Lemma~\ref{lem:weights}:
\beq
	\td{W}(X)= e^{-\mu |X|}\delta^{d(X)}\left[\prod_{r=1}^{\frac{L}{2}}|f(r)|^{d_r(X)} \right]\ . 
\eeq

\begin{lemma}
\label{lem:comb}
There exists a $L$- and $\beta$-independent constant $K_0>0$ of order one such that, if $\mu-3\delta c > K_0$, then
\beq
	q = \sum_{\chi\ni v}\td{W}(\chi)e^{14|\chi|} < 1\ ,
\eeq
where the sum is taken over all weakly-connected support sets $\chi$ that contain a fixed box or plaquette $v$.
\end{lemma}

To understand the physical meaning of this lemma, it is useful to interpret $\tilde{W}(\chi)$ as a Boltzmann weight of a classical statistical mechanics model in two dimensions.
Then the sum  $q = \sum_{\chi\ni v}\td{W}(\chi)e^{14|\chi|}$ can be thought of as a sum over all the Boltzmann weights of configurations containing a fixed point $v$. The condition that $q < 1$ is then analogous to the standard energy-entropy condition of Peierls~\cite{peierls} (see Refs.~\onlinecite{griffiths,thouless,landau-lifschitz} for related results). Specifically, in the expression $q = \sum_{\chi\ni v}\td{W}(\chi)e^{14|\chi|}$, the weight $\td{W}(\chi)$ can be thought of as a Boltzmann weight or energy contribution, while the
sum over weakly-connected support sets $\chi$ is the entropy contribution.\footnote{The exact
physical meaning of the exponential factor $e^{14|\chi|}$ is not as clear, but one interpretation
for this factor is given in Ref.~\onlinecite{brydges}.} The above lemma effectively shows that energy dominates over entropy, which tells us that the polymers do not proliferate and hence the perturbed model should be in the same phase as the unperturbed model.

Turning back to the proof: by combining Lemma~\ref{lem:weights} and Lemma~\ref{lem:comb} with standard results about the polymer expansion, it is straightforward to show that the polymer expansion 
for $\ln(Z_{\pm}/Z_0)$ is absolutely convergent in the region of $\mathbb{C}$
defined by $|\lambda|< \lambda_0$ for some $\lambda_0$ that is independent of $L$ and $\beta$.
With this convergence result in hand, we can apply the same reasoning as in KT (see pg. 470 of 
Ref.~\onlinecite{KT}) to conclude that,
for real $\lambda$ satisfying $|\lambda|< \lambda_0$, $H$ has
a unique ground state and a finite energy gap in each sector $\mathcal{H}_{\pm}$. This completes the
proof of part (1) of Theorem~\ref{thm:stability}.
In the rest of the second step of the proof we use the existence of the absolutely convergent polymer 
expansion for $\ln(Z_{\pm}/Z_0)$ to deduce several pieces of information about our model. 
Specifically, the polymer expansion allows us to prove the following two claims.

\begin{claim} \label{claim1}
For all $\lambda \in\mathbb{C}$ satisfying $|\lambda|< \lambda_0$, the complexified free energy
$-\ln(Z_{\pm})/\beta$ is a \emph{holomorphic} function of $\lambda$. In addition, $-\ln(Z_{\pm})/\beta$
is bounded from above as
\beq
	\Bigg|-\frac{1}{\beta}\ln\left(\frac{Z_{\pm}}{Z_{0}}\right) \Bigg| \leq c_3 L\ ,
\eeq
where the constant $c_3$ depends on $\Delta$ but not on $\lambda$, $h$, $\beta$, and $L$.
\end{claim}

The first part of this claim states that the free energy of our original Hermitian model (with
real $\lambda$) possesses an \emph{analytic continuation} to the region of 
$\mathbb{C}$ defined by $|\lambda|< \lambda_0$. The second part states
that the complexified free energy at any temperature is at most extensive in the 
system size. This result is not a surprise for real values of $\lambda$, 
but it is a nontrivial
result for complex $\lambda$. In particular, for complex $\lambda$ it is possible for $Z_{\pm}$ to be
equal to zero for some choice of parameters, and in that case the logarithm 
of $Z_{\pm}$ would diverge and be ill-defined at those parameter values. 
Claim~\ref{claim1} proves that this scenario cannot occur for $|\lambda|< \lambda_0$.

\begin{claim} \label{claim2}
Let $f(\lambda,\beta)$ be the difference between the complexified free energies for the two sectors 
$\mathcal{H}_{\pm}$,
\beq
	f(\lambda,\beta) = -\frac{1}{\beta}\ln\left(Z_{+}\right) + \frac{1}{\beta}\ln\left(Z_{-}\right) \ .
\eeq
Then the derivatives $\frac{d^n}{d \lambda^n}f(\lambda,\beta)$ 
vanish at $\lambda = 0$ for all $n$ satisfying $n < L/2$.
\end{claim}

This claim follows from the fact that 
one must flip all $L$ spins on the chain to go from the state with all spins up to the state
with all spins down. This result is not surprising and is clear from ordinary Schrodinger
perturbation theory. 

Claims~\ref{claim1} and \ref{claim2} are the key results from the second step of the proof. In the rest of this outline we explain how to finish the proof of Theorem~\ref{thm:stability} using these claims. This last part of the proof is a formalization
of the ideas of Klich~\cite{klich2010} and relies on a few important theorems from complex analysis.
To start, let $\mathcal{E}_{\pm}(\lambda)$ denote the zero temperature limit of the complexified
free energy,
\beq
	\mathcal{E}_{\pm}(\lambda) := \lim_{\beta\to\infty} -\frac{1}{\beta}\ln\left(Z_{\pm}\right)\ .
\eeq
For real values of $\lambda$ this limit clearly 
exists and is equal to the ground state energy $E_{\pm}(\lambda)$ of our model. 
Then, by the \emph{Vitali convergence theorem} (Theorem B.25 of Ref.~\onlinecite{FV}), this fact (convergence
for real $\lambda$), combined with Claim~\ref{claim1}, implies that the limit exists and that 
$\mathcal{E}_{\pm}(\lambda)$ is a holomorphic function of $\lambda$ for all $\lambda\in\mathbb{C}$ 
satisfying $|\lambda| < \lambda_0$.
In addition, since $\lim_{\beta\to \infty}Z_0 = 1$, and since the constant $c_3$ from Claim~\ref{claim1} 
was independent of $\beta$, the bound from Claim~\ref{claim1} carries over to 
$\mathcal{E}_{\pm}(\lambda)$ and we have
\beq
	\big|\mathcal{E}_{\pm}(\lambda)\big| \leq c_3 L\ .
	\label{claim2v2}
\eeq

We now show that these properties imply an exponential bound on the
energy difference $E_{+}(\lambda) - E_{-}(\lambda)$. Our strategy is to show that the
\emph{complexified} energy difference 
\beq
	f(\lambda) := \mathcal{E}_{+}(\lambda) - \mathcal{E}_{-}(\lambda)
\eeq
satisfies an exponential bound throughout the entire convergence region inside $\mathbb{C}$. This result
then immediately implies an exponential bound on $E_{+}(\lambda) - E_{-}(\lambda)$ for real values of 
$\lambda$ within the convergence region.

To show that $f(\lambda)$ satisfies an exponential bound we first note that $f(\lambda)$ is a holomorphic function of $\lambda$ for $|\lambda| < \lambda_0$ since $\mathcal{E}_{\pm}(\lambda)$ are holomorphic in this region. 
In addition, (\ref{claim2v2}) guarantees
that $|f(\lambda)| \leq 2 c_3 L$, while Claim~\ref{claim2} implies that
\beq
	\frac{d^n f(\lambda)}{d \lambda^n}\Big|_{\lambda=0} = 0 \ \ ,\quad n\ \in\ \left\{0,1,\dots,\tfrac{L}{2}-1\right\}\ . 
\eeq
With these three properties in hand, we can now use a version of 
\emph{Schwarz's lemma}\footnote{The usual version of Schwarz's lemma states that if
$f(z)$ is holomorphic for $|z| < z_0$, satisfies $f(0)=0$, and is bounded as
$|f(z)|\leq f_0$ for all $|z|< z_0$, then $|f(z)|\leq \frac{|z|}{z_0} f_0$ for all $|z|< z_0$.
To prove this one applies the maximum modulus principle to the function $\frac{f(z)}{z}$, which
is also holomorphic for $|z|< z_0$ by virtue of the fact that $f(0)=0$. To prove the 
version of Schwarz's lemma that we use here, one should instead apply the maximum
modulus principle to the function $\frac{f(z)}{z^{\frac{L}{2}}}$, which is holomorphic on 
$|z|< z_0$ if $f(z)$ \emph{and} the first $L/2-1$ derivatives of $f(z)$ vanish at $z=0$.} 
to conclude that, for $|\lambda| < \lambda_0$, $f(\lambda)$ satisfies the bound
\beq
	|f(\lambda)| \leq \left(\frac{|\lambda|}{\lambda_0} \right)^{\frac{L}{2}} f_0\ \ ,\quad f_0 = 2 c_3 L\ . 
\eeq
In particular, for real $\lambda$ the energy difference $|E_{+}(\lambda) - E_{-}(\lambda)|$ obeys
a bound of the form \eqref{eq:exponential-bound} with $c_1 = 2c_3$ and 
$c_2 = \frac{1}{2}\ln(\lambda_0/|\lambda|)$.

\subsection{Step 1: A formula for $Z_{\pm}$}
\label{sec:step1}
In this section we present the first step of our proof. In particular, we show how to derive
the expression \eqref{eq:Z-as-sum} for  
$Z_{\pm}/Z_0$ and the properties of the support sets $X$ and weights $W_{\pm}(X)$ that are
stated in Eqs.~\eqref{eq:config-factors}, \eqref{eq:weight-factors}, and Lemma~\ref{lem:weights}.
Since this construction largely follows KT~\cite{KT}, we only outline the main
steps here. We provide all of the details for this step in Appendix~\ref{app:weights}.

\subsubsection{Setting up the expansion of $Z_{\pm}$}
\label{sec:setup}

To set up the expansion of $Z_{\pm}$ we first expand $e^{-\beta H}$ in a Dyson series by iterating 
the \emph{Duhamel} 
formula,
\beq
	e^{-\beta (H_0+\lambda V)} = e^{-\beta H_0} - \lambda \int_0^{\beta}d\tau\ e^{-(\beta-\tau)H_0}V e^{-\tau(H_0+\lambda V)}\ . \label{eq:duhamel}
\eeq
While KT~\cite{KT} used the Trotter product formula for this step, the Duhamel formula was used in this way in 
Refs.~\onlinecite{borgs1996,datta1996}. If we iterate the Duhamel formula then we end up with
a series expansion for $e^{-\beta H}$ of the form
\beq
	e^{-\beta H}= e^{-\beta H_0} +\sum_{n=1}^{\infty} 
	(-\lambda )^n \int_0^{\beta} d\tau_n\ \cdots\int_0^{\tau_2}d\tau_1\ e^{-\beta H_0}V(\tau_n)\cdots V(\tau_1)\ ,
\eeq
where the integration in the $n$th term is over the region of $\mathbb{R}^n$ defined by 
$0\leq \tau_1 \leq \tau_2 \leq \cdots \leq \tau_n \leq \beta$, and where
\beq
	V(\tau)= e^{\tau H_0}V e^{-\tau H_0}\ .
\eeq

One important fact about this expansion is that it is absolutely convergent for a lattice 
system with a finite lattice size $L$ and a finite-dimensional Hilbert space on each
site. Indeed, we have the bound\footnote{To derive this
bound one should unpack the $V(\tau_k)$'s and note that $|| e^{-(\tau_{k+1} -\tau_k)H_0}||\leq 1$ 
because $H_0$ has a positive spectrum.}
\begin{widetext}
\beq
	\sum_{n\geq 0} \Big|\Big| (-\lambda )^n \int_0^{\beta} d\tau_n\ \cdots\int_0^{\tau_2}d\tau_1\ e^{-\beta H_0}V(\tau_n)\cdots V(\tau_1)\Big|\Big|  \leq e^{\beta|\lambda|\cdot||V||} < \infty\ ,
\eeq
where $||\mathcal{O}||$ denotes the usual operator norm of $\mathcal{O}$ (the 
largest singular value of $\mathcal{O}$). In addition, one can see from the derivation that
this bound continues to hold for complex values of $\lambda$.
\end{widetext}

The next step is to rewrite $V$ as a sum over terms $V_Y$ that act on
(unordered) subsets $Y\subset\{1,\dots,L\}$ of lattice sites,
\beq
	V= \sum_{Y}V_Y\ .
\eeq
For our model there are two types of subsets that contribute non-zero terms to $V$. 
First, if $Y=\{j\}$ contains the single site $j$, 
\beq
	V_{\{j\}}= h \sigma^x_j\ .
\eeq
Next, if $Y=\{j,k\}$ is a set of two 
distinct lattice sites $j$ and $k$, then we have
\beq
	V_{\{j,k\}}= f(|j-k|_{\text{p}})\sigma^x_j \sigma^x_k\ .
\eeq
With this notation the $n$th term in the expansion of $e^{-\beta H}$ now takes the form
\beq
	(-\lambda)^n \sum_{Y_1,\dots,Y_n}\int_0^{\beta} d\tau_n\ \cdots\int_0^{\tau_2}d\tau_1\ e^{-\beta H_0}V_{Y_n}(\tau_n)\cdots V_{Y_1}(\tau_1)\ .
\eeq

Finally, to evaluate $Z_{\pm}$ we need to trace over the sector $\mathcal{H}_{\pm}$ of the full
Hilbert space. We choose to evaluate this trace in a basis of a simultaneous eigenstates of 
$H_0$ and the Ising symmetry operator $\mathcal{S}$. To define this basis, 
let $s=(s^{(2)},\dots,s^{(L)})$ be an $(L-1)$-tuple of spin values,
with $s^{(j)}\in\{\up,\down\}$ for
$j\in\{2,\dots,L\}$. Here, $s^{(j)}$ labels the $z$-projection of the spin on site
$j$. Using these values, we first define the set of $2^{L-1}$ states $|s\ran$ via
\beq
	|s\ran= |\up,s^{(2)},\dots,s^{(L)}\ran\ .
\eeq
These states have the spin at site $1$ pointing up, and then the configurations of the rest of the
spins are specified by $s^{(2)},\dots,s^{(L)}$. Using the states $|s\ran$, we can then construct
an orthonormal basis of states $|s,\pm\ran$ for $\mathcal{H}_{\pm}$ via 
\beq
	|s,\pm\ran= \frac{1}{\sqrt{2}}\left(1\pm\mathcal{S}\right)|s\ran\ .
\eeq
As we mentioned above, the states $|s,\pm\ran$ are also eigenstates of the
unperturbed Hamiltonian $H_0$,  
\beq
	H_0|s,\pm\ran= \ep_s |s,\pm\ran\ ,
\eeq
where the energy $\ep_s$ is given by
\beq
	\ep_s = \Delta\times(\text{ number of domain walls in }|s\ran\ )\  \label{eq:energies}
\eeq
Note that for the unperturbed Hamiltonian $H_0$, the
eigenvalues $\ep_s$ do not depend on the Ising symmetry sector, and so all eigenstates of
$H_0$ are (at least) doubly-degenerate. This exact degeneracy will be split once we turn on the 
interaction $V$. 

For any operator $\mathcal{O}$, we can then compute $\text{Tr}_{\pm}\{\mathcal{O}\}$ using this
basis in the standard way: $\text{Tr}_{\pm}\{\mathcal{O}\} = \sum_s\lan s,\pm|\mathcal{O}|s,\pm\ran$,
where we sum over all $(L-1)$-tuples $s=(s^{(2)},\dots,s^{(L)})$. Therefore, at this point our
expansion for $Z_{\pm}$ takes the form
\beq
	Z_{\pm} = Z_{0} + \sum_{n=1}^{\infty}(-\lambda)^n \sum_{Y_1,\dots,Y_n}\int_0^{\beta} d\tau_n\ \cdots\int_0^{\tau_2}d\tau_1\ \sum_s \lan s,\pm| e^{-\beta H_0}V_{Y_n}(\tau_n)\cdots V_{Y_1}(\tau_1)|s,\pm\ran\ .
\eeq

The next step in obtaining Eq.~\eqref{eq:Z-as-sum} is
to follow KT~\cite{KT} and introduce a ``blocking'' in the (imaginary) time direction:
we divide the time interval 
$[0,\beta)$ into subintervals of length $\tau$, where $\tau$ is a lattice spacing in the
time direction, and we assume
that $\beta/\tau$ is an integer that we call $M$, i.e. $\frac{\beta}{\tau} = M \in \mathbb{N}$.

Using this blocking in the time direction, we can obtain Eq.~\eqref{eq:Z-as-sum} via
the following steps. We first show that each term in the Duhamel expansion of $Z_{\pm}$ has 
a representation in terms of a microscopic configuration $\mathcal{C}$ on the region 
$[1,L+1)\times[0,\beta)$. The configuration $\mathcal{C}$ will consist of the worldlines of 
domain walls in the spin chain. Next, each microscopic configuration $\mathcal{C}$ can in turn be assigned
a \emph{support set} $s(\mathcal{C})$, which is a geometric configuration on the blocked spacetime
lattice defined by the temporal lattice spacing $\tau$. By collecting all microscopic terms
with the same support set $s(\mathcal{C}) = X$, we can rewrite $Z_{\pm}/Z_{0}$ in the form
\eqref{eq:Z-as-sum} where, roughly speaking, we have
\beq   
	W_{\pm}(X) = \sum_{\mathcal{C};\ s(\mathcal{C}) = X}W_{\pm}(\mathcal{C})\ , \label{eq:sloppy-weight-def}
\eeq
and where $W_{\pm}(\mathcal{C})$ is the term in the Duhamel expansion that
corresponds to the microscopic configuration $\mathcal{C}$. Strictly speaking, this formula for $W_{\pm}(X)$ is not quite correct because the Duhamel expansion for $Z_\pm$ involves both discrete summation as well as time integration. In Appendix~\ref{app:weights}, we present a precise formula for $W_{\pm}(X)$ that takes care of this notational issue.

\subsubsection{Definition of the support sets $X$ and factorization of the weights}
\label{sec:support-sets}

We now define the microscopic configurations $\mathcal{C}$ and
the support sets $X$ in our model, and then we discuss the factorization properties of the
weights $W_{\pm}(X)$. To define the microscopic configurations we
study the integrand of a typical term at $n$th order in our
expansion of $Z_{\pm}$, for example the term 
\beq
	\sum_{s}\lan s,\pm| e^{-\beta H_0}V_{Y_n}(\tau_n)\cdots V_{Y_1}(\tau_1)|s,\pm\ran \nnb \ .
\eeq
Each matrix element 
\beq
	\lan s,\pm| e^{-\beta H_0}V_{Y_n}(\tau_n)\cdots V_{Y_1}(\tau_1)|s,\pm\ran \label{eq:term}
\eeq
in this sum is associated with a microscopic configuration 
$\mathcal{C}$.
To define $\mathcal{C}$, we insert a complete set of states
for $\mathcal{H}_{\pm}$ between every perturbation term in \eqref{eq:term} to obtain an expression
of the form
\begin{align}
\sum_{s_1,\dots,s_{n-1}} \lan s,\pm |V_{Y_n}|s_{n-1},\pm\ran\cdots\lan s_1,\pm|V_{Y_1}|s,\pm\ran\ e^{-(\beta-\tau_n+\tau_1)\ep_{s}}e^{-(\tau_n-\tau_{n-1})\ep_{s_{n-1}}}\cdots e^{-(\tau_2-\tau_1)\ep_{s_1}} \nnb\ .
\end{align}
Since each perturbation term $V_Y$ flips either one or two spins and does nothing else, 
the summand here is only non-zero for a single choice of the intermediate states 
$|s_1,\pm\ran,\dots,|s_{n-1},\pm\ran$. If we denote these particular states by 
$|s'_1,\pm\ran,\dots,|s'_{n-1},\pm\ran$, then we 
find that \eqref{eq:term} now takes the form
\beq
	\lan s,\pm |V_{Y_n}|s'_{n-1},\pm\ran\cdots\lan s'_1,\pm|V_{Y_1}|s,\pm\ran\ e^{-(\beta-\tau_n+\tau_1)\ep_{s}}e^{-(\tau_n-\tau_{n-1})\ep_{s'_{n-1}}}\cdots e^{-(\tau_2-\tau_1)\ep_{s'_1}}\ . 
	\label{eq:matrix-element-2}
\eeq

We can associate \eqref{eq:matrix-element-2} with a microscopic configuration
$\mathcal{C}$ of domain wall worldlines on $[1,L+1)\times[0,\beta)$. To do this, we first draw vertical lines in the interval $[\tau_1,\tau_2]$ at the horizontal 
locations\footnote{The horizontal location of a domain wall is the midpoint of the horizontal bond
that lies between the two oppositely oriented
spins that form that domain wall.} of all domain walls in $|s'_1,\pm\ran$, then draw
vertical lines in the interval $[\tau_2,\tau_3]$ at the horizontal locations of 
all domain walls in $|s'_2,\pm\ran$, and so on. The last time interval wraps around the
time direction (the trace enforces periodic boundary conditions in time), 
and so in that case we draw the vertical lines for
domain walls in $|s,\pm\ran$ in the interval
$[\tau_n,\beta]$ and then continue them in the interval $[0,\tau_1]$. At this point
the configuration $\mathcal{C}$ consists of vertical segments, but has no horizontal segments.
It turns out that this is enough information for our derivation, and we do not need
to assign any horizontal segments to $\mathcal{C}$.

For every microscopic configuration $\mathcal{C}$, we define a corresponding support
set $X = s(\mathcal{C})$.
Each support set is constructed from \emph{plaquettes}, 
\emph{boxes}, and \emph{dashed lines} on the blocked spacetime lattice. The
plaquettes and boxes were already considered in KT~\cite{KT}, but the dashed lines
are a new ingredient that is necessary to keep track of the long-range interactions in our model. 
The plaquettes can be viewed as subsets of $\mathbb{R}^2$ and they take the form  
$[j,j+1]\times[\tau(\ell-1),\tau\ell]$, where $j\in\{1,\dots,L\}$ and 
$\ell\in\{1,\dots,M\}$. The
boxes can also be viewed as subsets of $\mathbb{R}^2$, and they take the form 
$[j-\frac{1}{2},j+\frac{1}{2}]\times[\tau(\ell-1),\tau\ell]$ for
$j\in\{1,\dots,L\}$ and $\ell\in\{1,\dots,M\}$. [In these definitions, horizontal coordinates
should always be interpreted modulo $L$.] Finally, each dashed line in a support
set $X$ connects two different boxes or two different plaquettes in the same time slice of $X$.\footnote{As we will see below, support sets that contain dashed lines connecting plaquettes to plaquettes have vanishing weight and do not appear in our expression for $Z_\pm$. The reason we include these ``illegal'' configurations is to simplify the structure of the support sets by putting boxes and plaquettes on an equal footing.}As such, dashed lines are always
parallel to the spatial direction of the spacetime lattice. Finally, our rule for assigning dashed
lines to a support set $X$ will be such that any two boxes
within a given time slice are connected by at most one dashed line.

We now present the rules for determining which plaquettes, boxes, and dashed lines are
included in the support set $s(\mathcal{C})$ for the configuration $\mathcal{C}$
associated with a particular matrix element like \eqref{eq:term}. 
If $\mathcal{C}$ has a worldline that passes all the way 
through a given plaquette, then $s(\mathcal{C})$ contains
this plaquette. For this to occur, $\mathcal{C}$ must contain
an unbroken vertical segment of length at least $\tau$. Next, 
if \eqref{eq:term} contains a perturbation term $V_Y$ with support on site $j$ and acting
within the time interval $[\tau(\ell-1),\tau\ell]$, then $s(\mathcal{C})$ 
contains the box centered on $j$ within this time slice. With these
rules, the worldlines in $\mathcal{C}$ will always begin and end inside boxes that
are part of $s(\mathcal{C})$.

\begin{figure}[t]
  \centering
    \includegraphics[width= 0.75\textwidth]{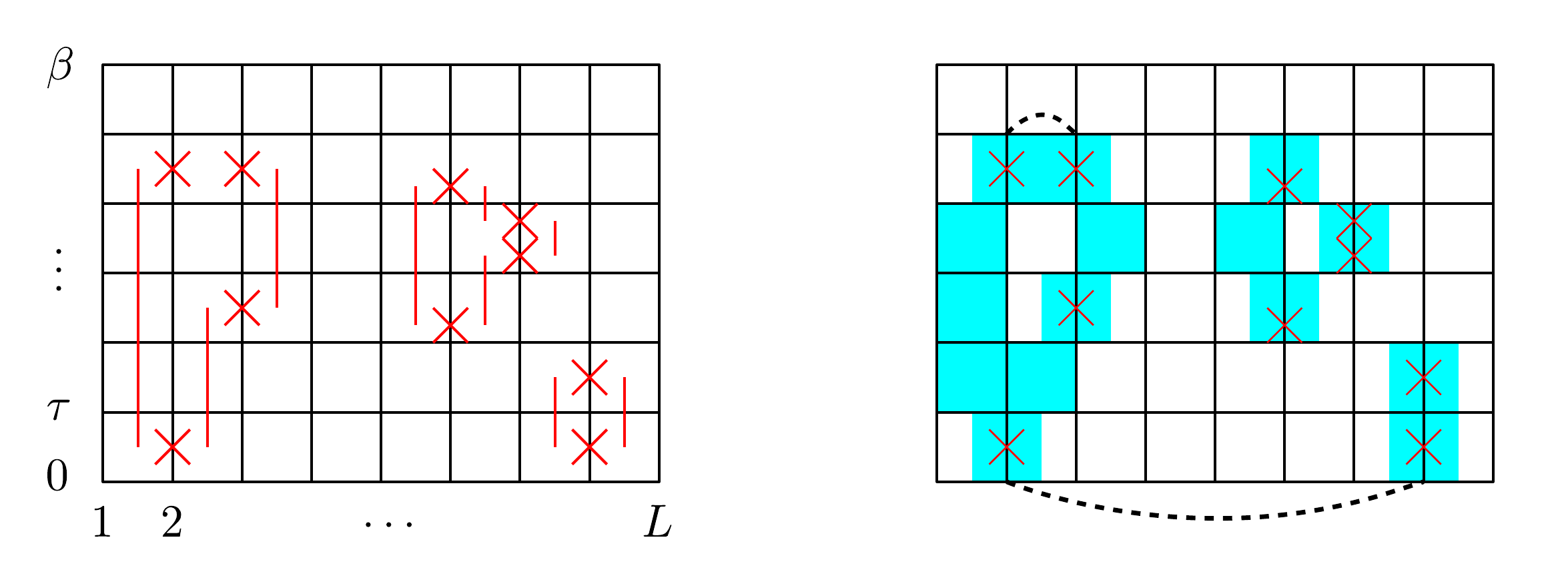} 
\caption{A microscopic domain wall worldline configuration (the red lines
in the left panel) and its corresponding support set (shown in the right panel). The support set is made
up of boxes, plaquettes, and dashed lines, and the plaquettes and boxes are shaded in light blue. The red 
crosses indicate the action of a $\sigma^x$ operator at the given position in spacetime.}
\label{fig:support-sets}
\end{figure}

We now come to the rule for assigning dashed lines to $s(\mathcal{C})$.
If \eqref{eq:term} contains a perturbation $V_{\{j,k\}}$ (i.e., $V_Y$ with $Y=\{j,k\}$ and $j\neq k$) 
acting within the time interval $[\tau(\ell-1),\tau\ell]$, 
then $s(\mathcal{C})$ contains a \emph{single} dashed line connecting the boxes centered on $j$ 
and $k$ within this time slice (those boxes will already be in $s(\mathcal{C})$
because of the rule for assigning boxes to a support set). The dashed line that
we draw to connect the two boxes is always the dashed line of minimal length according to the
periodic distance function $|\cdot|_{\text{p}}$. Therefore, the maximum length that a dashed line
can have is $L/2$. In Fig.~\ref{fig:support-sets} we show an example of a microscopic domain wall worldline 
configuration (left panel) and its corresponding support set (right panel).

For a given support set $X$, we define $p(X)$ and $b(X)$ 
to be the number of plaquettes and boxes, respectively, in $X$. Next, we define
$d_r(X)$ to be the number of dashed lines of length $r$ in $X$, where $r\in\{1,\dots,\tfrac{L}{2}\}$. The
total number of dashed lines in $X$ is then given by $d(X):= \sum_{r=1}^{\frac{L}{2}}d_r(X)$. 
Finally, we define the ``size'' $|X|$ of the support set 
$X$ to be equal to the total number of plaquettes and boxes in $X$, 
\beq
	|X|:= p(X) + b(X)\ .
\eeq
Note that we \emph{do not} include the number of dashed lines in the definition of $|X|$.

To proceed further, we define two different notions of 
connected support sets which will be important in our proof. We call a support set $X$ \emph{connected} 
if (i) $X$ has no dashed lines, and (ii) the boxes and plaquettes in $X$ form a connected subset of 
$\mathbb{R}^2$. Note that in this definition, two boxes that only touch at a corner form
a connected subset of $\mathbb{R}^2$ since we have defined the boxes to be closed on all sides. The same goes for two plaquettes that only touch at a corner. Thus, each box or plaquette is touching 14 neighboring boxes and/or plaquettes.
We use the special notation $\td{\chi}$ to denote a connected support set.

Likewise, we call a support set $X$ \emph{weakly-connected} if it is possible to get from any connected component of $X$ (regarded as a subset of $\mathbb{R}^2$) to any other connected component of $X$ by traversing a finite sequence of dashed lines belonging to $X$. We use the notation $\chi$ (no tilde) to denote a weakly-connected support set.

We also need to define a notion of \emph{intersection} for 
weakly-connected support sets. We say that two weakly-connected support sets $\chi_1$ and $\chi_2$ 
intersect, denoted by 
\beq
	\chi_1 \cap \chi_2 \neq \varnothing\ , \nnb
\eeq
if $\chi_1$ and $\chi_2$ share a box or plaquette or if there is a box or plaquette in $\chi_1$ that is touching a box or plaquette
in $\chi_2$ (including touching at corners as we discussed above). 
The right panel of Fig.~\ref{fig:support-sets} features one weakly-connected support set and one
connected support set, and these two do not intersect with each other.

With these definitions it is clear that any support set $X$ can be expressed as a
collection of a finite number of \emph{non-intersecting} 
weakly-connected support sets, 
\beq
	X=\chi_1 \cup \chi_2 \cup \cdots \ \ , \quad \quad \chi_i \cap\chi_j = \varnothing\ \ \forall\ i, j\ .
\eeq
Furthermore, it is not hard to see that the weights $W_{\pm}(X)$ factor into a product of 
weights for each weakly-connected support set in $X$,
\beq
	W_{\pm}(X)= \prod_{i=1}^k W_{\pm}(\chi_i)\ .
\eeq

The proof of this factorization property for our model is identical to the proof of the
corresponding property in KT~\cite{KT} (see their Lemma 4.5), and so we do not repeat it here. It follows
from the fact that the weights $W_{\pm}(\mathcal{C})$ for the microscopic configurations 
$\mathcal{C}$ have this factorization property, and this is because 
$W_{\pm}(\mathcal{C})$ is proportional to the matrix element \eqref{eq:matrix-element-2}.

\subsubsection{The bound on the weights}

The last result from step one of the proof is 
Lemma~\ref{lem:weights}, which shows that the weights $W_{\pm}(X)$ are bounded by
\beq
	|W_{\pm}(X)| \leq e^{-\mu |X|}\delta^{d(X)}\left[\prod_{r=1}^{\frac{L}{2}}|f(r)|^{d_r(X)} \right] 
\eeq
for any $\delta, \mu > 0$, as long as $|\lambda|$ is smaller than some $\lambda_0$. Recall that in this bound, 
$|X|$ is the number of boxes and plaquettes in $X$, $d_r(X)$ is the number of dashed lines of length 
$r$ in $X$, and $d(X)=\sum_{r=1}^{\frac{L}{2}}d_r(X)$ is the total number of dashed lines in $X$.
This bound will be important for the second step of our proof, where we use it as part of the proof that the 
polymer expansion for our model is absolutely convergent. We present a detailed proof of Lemma~\ref{lem:weights} in Appendix~\ref{app:weights}.

\subsection{Review of the polymer expansion}
\label{sec:polymer}

Before moving on to the second step of the proof, we first introduce the main tool that is used in
that step, namely the \emph{polymer
expansion} (also known as the \emph{cluster expansion}). 
The polymer expansion is a tool from rigorous statistical 
mechanics that, under certain conditions, allows one to obtain an absolutely convergent series expansion for 
the logarithm of the partition function of a given model. One can then prove
many things about the model using this expansion. In this subsection we give a 
brief overview of the polymer expansion and the conditions that guarantee its convergence
(we discuss the issue of convergence in more detail in Appendix~\ref{app:polymer}). For our
presentation of these results we follow Ref.~\onlinecite{brydges}, Ch.~5 of Ref.~\onlinecite{FV}, and
Ch.~20 of Ref.~\onlinecite{GJ}. 

In practice, the polymer expansion arises as follows. In many statistical mechanical models,
the partition function $Z$ admits an absolutely convergent 
series expansion in which each term in the series has a geometric interpretation in terms of
\emph{disconnected} geometric objects. These geometric objects are the polymers that give
this expansion its name.
A good example to have in mind is the contours of domain walls that appear in the low temperature
expansion of the classical Ising model in two dimensions. 
When this geometric expansion for $Z$ exists, one can sometimes
find a region of the model's parameter space in which the logarithm $\ln(Z)$ of the partition function also
admits an absolutely convergent series expansion in terms of the same geometric objects. This 
expansion for $\ln(Z)$ is the polymer expansion.

Let us now be more explicit. Let $\Gamma$ be a finite set whose elements are
called \emph{polymers} and are denoted by $\gamma$.
The polymers have a notion of intersection so that, for any two polymers $\gamma_1$ and 
$\gamma_2$, we can have $\gamma_1\cap\gamma_2 = \varnothing$ or 
$\gamma_1\cap\gamma_2\neq \varnothing$. Note
that the intersection of a polymer with itself is non-empty, 
$\gamma\cap\gamma\neq \varnothing$.
To define a statistical model we also need to define the \emph{weight} of a polymer. 
The weight $W(\gamma)\in\mathbb{C}$ of a polymer $\gamma$ is just
a complex number associated with that polymer.

Let $\Gamma'$ be a subset of $\Gamma$. It is an unordered collection of a certain number of 
polymers. A subset $\Gamma'$ of polymers of the form $\Gamma'= \{\gamma_1,\dots,\gamma_k\}$ 
is said to be \emph{disconnected} if the polymers $\gamma_1,\dots,\gamma_k$ are pairwise disjoint,
i.e., $\gamma_i \cap \gamma_j = \varnothing$ for $1\leq i < j \leq k$. If
$\Gamma'= \{\gamma_1,\dots,\gamma_k\}$ is disconnected, then we define
the weight of $\Gamma'$ to be the product of the weights of all the $\gamma_i$, 
\beq
	W(\Gamma') := \prod_{i=1}^k W(\gamma_i) \ .
\eeq
In terms of these quantities, the \emph{polymer partition function} $Z$ is defined as
\beq
	Z:=\sum_{\substack{\Gamma'\subseteq\Gamma\\ \text{disconnected}}}W(\Gamma')\ , \label{eq:Z-polymer}
\eeq
where the sum is taken only over disconnected subsets $\Gamma'$. In this sum we also include
the empty set $\Gamma'=\varnothing$, and we assign it a weight of $1$,  $W(\varnothing):= 1$.
Note also that, although we have called $Z$ the polymer ``partition function'', we have allowed
the weights $W(\gamma)$ to be complex numbers. As a result, the polymer expansion can be applied to both
quantum and classical statistical mechanics.

The polymer expansion is an expansion for $\ln(Z)$ in terms of the same weights
$W(\gamma)$ that appear in the definition of $Z$. To write down the series for 
$\ln(Z)$, we first need to introduce some more notation. If we are given a collection of polymers 
$\gamma_1,\dots,\gamma_k$, then we define $G(\gamma_1,\dots,\gamma_k)$ to be the graph 
with the following properties. First, the vertices of $G(\gamma_1,\dots,\gamma_k)$ are labeled from
$1$ to $k$. Second, $G$ has an edge between vertices $i$ and $j$ (with $i\neq j$) if
$\gamma_i\cap\gamma_j \neq\varnothing$ (for $i,j\in\{1,\dots,k\}$), and no other edges besides these. 
Next, we need to define the \emph{index} of a connected graph. Let $C_k$ be the set of
connected graphs with $k$ vertices labeled from $1$ to $k$, and let $G\in C_k$. The index of $G$, denoted by $n(G)$, is defined by
\beq
	n(G):= \sum_{\substack{H\in C_k\\ H\subseteq G}}(-1)^{\ell(H)}\ ,
\eeq
where the sum is taken over all $H$ in $C_k$ that are also subgraphs of $G$ (including
$H=G$), and where $\ell(H)$ is the number of edges (or lines) in $H$. 

Using these notations, the
polymer expansion for $\ln(Z)$ is given by
\begin{align}
	\ln(Z) &= \sum_{k=1}^{\infty}\frac{1}{k!}\sum_{H\in C_k}n(H)\sum_{\substack{\gamma_1,\dots,\gamma_k\\ G(\gamma_1,\dots,\gamma_k)=H}} \prod_{i=1}^k W(\gamma_i) \\
	&:= \sum_{k=1}^{\infty}I_k\ , \label{eq:polymer-expansion}
\end{align}
where we defined $I_k$ to be the $k$th term in the sum in the first line.
Note that $\ln(Z)$ only receives contributions from (ordered) collections $\gamma_1,\dots,\gamma_k$ of 
polymers whose associated graph $G(\gamma_1,\dots,\gamma_k)$ is connected.

We now review the conditions that ensure the convergence of the polymer expansion for $\ln(Z)$.
For this purpose we specialize to the concrete setting of polymer models defined on a
finite set $\Lambda$ whose elements are called ``vertices.'' 
In this setting each polymer $\gamma$ contains a
subset of the vertices in $\Lambda$. A polymer can also have additional internal structure beyond the 
vertices that it contains, and we will see an example of this when we apply the polymer expansion to our 
model. We define the size $|\gamma|$
of a polymer to be the number of vertices contained in $\gamma$ (so $|\gamma|\geq 1$), and the
total number of vertices in $\Lambda$ is denoted by $|\Lambda|$. Finally, the notion of intersection
is simply sharing of vertices (i.e., $\gamma_1 \cap \gamma_2 \neq \varnothing$ if $\gamma_1$
and $\gamma_2$ have at least one vertex in common).

We now present a simple convergence criterion for the polymer expansion in
this setting. (See Appendix~\ref{app:polymer} for a derivation of this criterion starting from the general convergence
criterion of Ref.~\onlinecite{brydges}). To use this criterion, our polymer model needs to have the following property: the weights need to satisfy an upper bound of the form 
\beq
	|W(\gamma)| \leq \td{W}(\gamma)\ , \label{eq:w-tilde}
\eeq
where the set of polymers $\Gamma$ is \emph{isotropic} with respect to the weight $\td{W}(\gamma)$, in the sense that the sum $\sum_{\gamma\ni v}\td{W}(\gamma)g(|\gamma|)$ is independent of the vertex $v \in \Lambda$ for any function $g(|\gamma|)$ of the polymer size $|\gamma|$ (the sum is taken over all $\gamma \in \Gamma$ containing $v$).

Let us assume that the above property holds. If we define
\beq
	q := \sum_{\gamma\ni v} \td{W}(\gamma)e^{|\gamma|}\ , \label{eq:q-def}
\eeq
then one can establish the following convergence criterion:

\begin{thm}[Simplified convergence criterion]
\label{thm:convergence}
If $q < 1$, then the polymer expansion \eqref{eq:polymer-expansion} for $\ln(Z)$ is absolutely convergent
for fixed $|\Lambda|$, and furthermore $\ln(Z)$ satisfies the bound
\beq
	|\ln(Z)|\leq \sum_{k=1}^{\infty}|I_k| \leq -|\Lambda|\ln(1-q)\ .
\eeq
\end{thm}

To complete our review of the polymer expansion, we need to discuss the properties of this expansion when the 
weights $W(\gamma)$ depend
on an additional parameter $z\in\mathbb{C}$, so that we have $W(\gamma)\to W_z(\gamma)$ in our 
original expansion for $Z$. Suppose that $W_z(\gamma)$ is a holomorphic function of 
$z$ for $z\in \mathcal{D}$, where $\mathcal{D}$ is a connected open subset of $\mathbb{C}$. 
In this situation it
is natural to ask whether $\ln(Z)$ is also a holomorphic function of $z$ on $\mathcal{D}$ when the polymer
expansion converges. The answer to this question is provided by Theorem 5.8 of Ref.~\onlinecite{FV}, which
can be summarized as follows.\footnote{Note that Ref.~\onlinecite{FV} refers
to a connected open subset of $\mathbb{C}$ as a \emph{domain}.} 
Suppose that $\sup_{z\in \mathcal{D}}|W_z(\gamma)| \leq \ov{W}(\gamma)$,
where $\ov{W}(\gamma)$ is a real and positive weight that is independent of $z$, and let
$\ov{Z}$ be the partition function obtained by replacing $W_z(\gamma)$ with $\ov{W}(\gamma)$
in Eq.~\eqref{eq:Z-polymer}.
If the polymer expansion for $\ln(\ov{Z})$ is absolutely convergent, 
then the polymer expansion for $\ln(Z)$ is absolutely convergent for any
$z\in \mathcal{D}$, and $\ln(Z)$ is a holomorphic function of $z$ for $z\in \mathcal{D}$. As
we discussed above, this holomorphic property of 
$\ln(Z)$ is a crucial ingredient in our stability proof.

\subsection{Step 2: Convergent polymer expansion and its consequences}
\label{sec:step2}

\subsubsection{Mapping our model to a polymer model}
\label{sec:graph-mapping}

We start by interpreting our formula \eqref{eq:Z-as-sum} for $Z_{\pm}/Z_0$ as a polymer partition function. To this end, recall that the support sets $X$ can be decomposed into a union of \emph{non-intersecting}
weakly-connected support sets $\chi_1,\chi_2,\dots$, and that
$W_{\pm}(X)$ factors according to this decomposition (\ref{eq:weight-factors}). It follows that the polymers in our model correspond exactly to the weakly-connected support sets $\chi$.

To simplify the notation, we now reformulate this polymer model so that each polymer is described by a subset of vertices in a particular lattice $\Lambda$. This lattice $\Lambda$ consists of spacetime points that are at the 
\emph{plaquette centers} and \emph{vertical bond 
midpoints} of the blocked spacetime lattice. Since the 
blocked spacetime lattice has sites at locations $(j,(\ell-1)\tau)$ for 
$j\in\{1,\dots,L\}$ and $\ell\in\{1,\dots,M\}$, the vertices in $\Lambda$ are located at
$(j+1/2,(\ell-1/2)\tau)$ (the plaquette centers) and $(j,(\ell-1/2)\tau)$ (the vertical bond
midpoints), for $j\in\{1,\dots,L\}$ and $\ell\in\{1,\dots,M\}$.
The total number of vertices in $\Lambda$ is then $|\Lambda|= 2 L M$ where $M$ is the integer $M=\beta/\tau$, defined previously.

We can map each weakly-connected support set $\chi$ to a polymer
$\gamma_{\chi}$ that consists of a subset of vertices of $\Lambda$ together with appropriate dashed lines connecting these vertices. The mapping that we use is the obvious one:
\begin{enumerate}
\item If $\chi$ contains a certain box or plaquette, then $\gamma_{\chi}$ contains the vertex at the center 
of that box or plaquette.
\item If two boxes or two plaquettes in $\chi$ are connected by a dashed line, then the corresponding 
vertices in $\gamma_{\chi}$ are also connected by a dashed line.
\end{enumerate}
We also define the weight of $\gamma_{\chi}$ to be equal to the 
weight of $\chi$, that is: $W_{\pm}(\gamma_{\chi}) := W_{\pm}(\chi)$. Finally, we use the special notation $\td{\gamma}_{\td{\chi}}$ to denote a polymer that is obtained from a connected support set $\td{\chi}$.

The price we pay for this simple mapping from $\chi$ to $\gamma_{\chi}$ is that our notion of
intersection of weakly-connected support sets does not quite map onto the simple notion of 
sharing of vertices of $\Lambda$ that we considered in Sec.~\ref{sec:polymer}. 
We now discuss the notion of intersection in our model in more detail, and also 
clarify the structure of the set $\Gamma$ of allowed polymers in our model.

\begin{figure}[t]
  \centering
    \includegraphics[width= .3\textwidth]{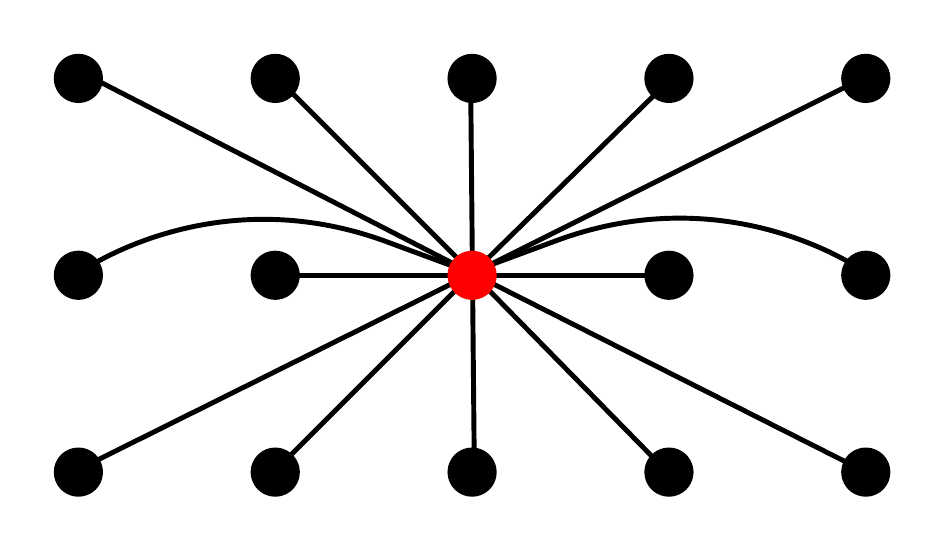} 
\caption{We endow $\Lambda$ with a graph structure such that each vertex is connected by edges
to 14 of its neighbors. This figure shows the 14 edges that emanate from the central red vertex
(which could represent a box or a plaquette on the blocked spacetime lattice). The 
edges in $\Lambda$ that do not connect to the red vertex are not shown in this figure.}
\label{fig:edges}
\end{figure}

To describe the set of allowed polymers and the notion of 
intersection, it is convenient to add additional structure to the set $\Lambda$ to turn it into a connected 
graph. We do this by adding
edges\footnote{It is very important to note that these edges are different from the dashed lines that we previously discussed.} that connect each vertex to 14 of its neighbors in the manner shown in Fig.~\ref{fig:edges}.
To understand these connections, recall that in our original picture on the blocked spacetime lattice,
each box touches 14 neighboring boxes and/or plaquettes, and likewise for each plaquette. The 14 connections
shown in Fig.~\ref{fig:edges} correspond exactly to these neighboring boxes/plaquettes.

Now that we have endowed $\Lambda$ with a graph structure, we are almost ready to describe the set of allowed polymers
$\Gamma$. To do this we first need to review the definition of an \emph{induced
subgraph} of a graph. Consider an abstract graph $G = (V,E)$, where $V$ and $E$ are the sets of vertices and edges of $G$. If $V'\subset V$ is a subset of vertices, then the 
induced subgraph of $G$ determined by $V'$ takes the form $(V',E')$, where $E' \subset E$ contains all edges 
of $G$ that have both of their endpoints contained in $V'$. In general, an induced subgraph will not 
be connected and will instead have several connected components. An induced subgraph that is connected
is called a \emph{connected induced subgraph}.

We are now ready to describe $\Gamma$. The set $\Gamma$ contains all polymers 
$\gamma$ with the following structure:
\begin{enumerate}
\item $\gamma$ contains a subset of the vertices in $\Lambda$. 
\item $\gamma$ can contain horizontal dashed lines connecting pairs of vertices with
the same time coordinate. The two vertices connected by the dashed line must come from the
same type of object (box or plaquette) on the blocked spacetime lattice.
\item Two vertices in $\gamma$ are connected by at most one dashed line.
\item Let $\Lambda_{\gamma}$ be the induced subgraph of $\Lambda$ determined by the vertices in $\gamma$. 
If $\Lambda_{\gamma}$ has more than one connected component, then the dashed lines in $\gamma$ must
be such that it is possible to get from one connected component of $\Lambda_{\gamma}$ to any other one by
traversing sufficiently many dashed lines.
\end{enumerate}
Although it looks different, this definition of the allowed polymers is exactly the same as the definition
of the allowed weakly-connected support sets that we presented in Sec.~\ref{sec:support-sets}. 
We have simply translated
that definition from the language of boxes and plaquettes into the language of polymers $\gamma$ containing 
vertices of the connected graph $\Lambda$. We also note that our definition of \emph{connected} support sets
maps onto special polymers $\td{\gamma}$ that (i) do not contain any dashed lines and (ii) whose induced subgraph $\Lambda_{\td{\gamma}}$ has only one connected 
component, i.e., $\Lambda_{\td{\gamma}}$ is a connected induced subgraph of $\Lambda$.

Finally, using the graph structure of $\Lambda$, we can now translate the notion of
intersection of weakly-connected support sets into the polymer language. Specifically, we find that
two polymers $\gamma_1$ and $\gamma_2$ intersect if (i) they have one or more vertices in common, or 
(ii) a vertex in $\gamma_1$ is connected to a vertex in
$\gamma_2$ via an edge of $\Lambda$.

\subsubsection{Lemma 2 and the convergence of the polymer expansion}
\label{sec:lemma2}

At this point we have succeeded in showing that our expansion for $Z_{\pm}/Z_0$ has exactly the
structure required to apply the polymer expansion to study $\ln(Z_{\pm}/Z_{0})$.
Now we just need to show that the resulting expansion is absolutely convergent. 

We would like to prove convergence using the convergence criterion that we
presented in Theorem~\ref{thm:convergence}. 
However, the notion of intersection in our model is slightly more complicated than the simple notion of 
intersection that was assumed in that theorem. It turns out that this issue is not hard to deal with, and we only need a minor modification of 
the criterion from Theorem~\ref{thm:convergence} to prove convergence in our 
case. Specifically, if we re-define $q$ as
\beq
	q := \sum_{\gamma\ni v}\td{W}(\gamma)e^{14|\gamma|}\ ,
\eeq
where the only change is the multiplicative factor of $14$ in the exponent, 
then our polymer expansion for $\ln(Z_{\pm}/Z_0)$ will be absolutely convergent if $q < 1$. We explain
the derivation of this modified convergence criterion in Appendix~\ref{app:polymer}.

As in Sec.~\ref{sec:polymer}, this
(modified) convergence criterion only holds if our polymer model satisfies the following property: the weights need to satisfy an upper bound of the form  
$|W_{\pm}(\gamma)| \leq \tilde{W}(\gamma)$ where the set of polymers $\Gamma$ is \emph{isotropic} with respect to the weight $\tilde{W}(\gamma)$ in the sense of Sec.~\ref{sec:polymer}. But we have already established exactly such a bound in 
step one of the proof. Indeed, Lemma~\ref{lem:weights} implies that 
the weight $W_{\pm}(\gamma)$ obeys the bound $|W_{\pm}(\gamma)| \leq \td{W}(\gamma)$ with 
\beq
	\td{W}(\gamma) = e^{-\mu |\gamma|}\delta^{d(\gamma)}\left[\prod_{r=1}^{\frac{L}{2}}|f(r)|^{d_r(\gamma)} \right]\ , \label{eq:tilde-weight}
\eeq
where $d_r(\gamma)$ is the number of dashed lines of length $r$ in $\gamma$ and 
$d(\gamma)$ is the total number of dashed lines in $\gamma$. Moreover, it is clear from the form of $\tilde{W}(\gamma)$, that our set of polymers $\Gamma$ is isotropic with respect to this weight.

At this point, all that remains to prove the convergence of the
polymer expansion is to show that $q < 1$. More precisely, we need to show that $q < 1$ for at least one choice of $\mu, \delta > 0$. Then, as a result of Lemma~\ref{lem:weights}, there exists $\lambda_0$ so that, for all $\lambda\in\mathbb{C}$ satisfying $|\lambda| < \lambda_0$, we have $|W_{\pm}(\gamma)|\leq \td{W}(\gamma)$, and hence the polymer expansion converges. 

The desired bound, $q < 1$, follows from Lemma~\ref{lem:comb}, which we prove in Appendix~\ref{app:combinatorics} using combinatorial arguments.

\subsubsection{Claims 1 and 2 as consequences of the convergent polymer expansion}

To wrap up our discussion of step two of the proof, we now explain how Claims~\ref{claim1} and \ref{claim2} can be
deduced from the existence of the convergent polymer expansion for $\ln(Z_{\pm}/Z_0)$. To start, 
the first part of Claim~\ref{claim1} 
follows from the general results about the holomorphic properties of the polymer expansion that
we reviewed in the last paragraph of Sec.~\ref{sec:polymer}. Recall from that paragraph that the 
key to proving holomorphicity in a connected open set $\mathcal{D}$ of the (complexified) parameter space 
is to show that the polymer expansion also converges if we replace the
weights $W_{\pm}(X)$ by a uniform upper bound $\ov{W}(X)$ that is independent of the parameters
and holds throughout the region $\mathcal{D}$. 
But we have already established this, since we can use for
$\ov{W}(X)$ the same upper bound $\td{W}(X)$ that we obtained in Lemma~\ref{lem:weights} and used
in Lemma~\ref{lem:comb}. Thus, we can conclude that the complexified free energy $-\ln(Z_{\pm})/\beta$
in our model is a holomorphic function of $\lambda$ in the region of $\mathbb{C}$
where the polymer expansion converges.  

Next, the second part of Claim~\ref{claim1} is also a simple consequence of the convergence of the polymer 
expansion. Since the expansion for $\ln(Z_{\pm}/Z_{0})$ converges absolutely, 
Theorem~\ref{thm:convergence} implies that $\ln(Z_{\pm}/Z_{0})$ obeys the bound
\beq
	\Bigg| \ln\left(\frac{Z_{\pm}}{Z_{0}}\right) \Bigg| \leq - |\Lambda|\ln(1-q)\ .
\eeq
[This bound has the same form as in Theorem~\ref{thm:convergence}, but $q$ here is now our modified 
version with the extra factor of 14 in the exponent.]
For our model the size of $\Lambda$ is given by $|\Lambda| = 2 L M$, where
$M = \beta/\tau$. Thus, Claim~\ref{claim2} holds with the constant $c_3$ given by
\beq
	c_3 = - \frac{2}{\tau}\ln(1-q)\ .
\eeq
Since $c_3$ can be expressed only in terms of $\tau$ and $q$, we find that $c_3$ depends only on 
$\Delta$, the constant $c$ from the summability condition \eqref{eq:sum-bound}, and geometric properties of 
the model.\footnote{Specifically, the combinatorial factors $K$ and $K_0$ that appear in our proofs
in Appendixes~\ref{app:weights} and \ref{app:combinatorics}.} Note that $c_3$ \emph{does not} depend on 
$\lambda,h,\beta$, or $L$. 

Finally, we explain how Claim~\ref{claim2} follows from the form of the polymer expansion for 
$\ln(Z_{\pm}/Z_{0})$. 
To understand this claim we first recall that for any support set $X$ the weight $W_{\pm}(X)$ is given by 
a combined sum and integral over terms of the form
\beq
	\frac{1}{Z_0}\text{Tr}_{\pm}\{e^{-\beta H_0}V_{Y_n}(\tau_n)\cdots V_{Y_1}(\tau_1)\}\ ,
\eeq
where $\tau_1,\dots,\tau_n$ and $Y_1,\dots,Y_n$ are a sequence of times and perturbation terms that
are consistent with the support set $X$. Now, for any operator $\mathcal{O}$, the difference between
the trace of $\mathcal{O}$ over $\mathcal{H}_{+}$ and the trace of $\mathcal{O}$ over $\mathcal{H}_{-}$
can be written in the form
\beqa
	\text{Tr}_{+}\{\mathcal{O}\} - \text{Tr}_{-}\{\mathcal{O}\} &=& \text{Tr}\{\mathcal{P}_{+}\mathcal{O}\} - \text{Tr}\{\mathcal{P}_{-}\mathcal{O}\} \nnb \\
	&=& \text{Tr}\{\mathcal{S}\mathcal{O}\}\ , \label{eq:trace-diff}
\eeqa
where $\mathcal{S}$ is the Ising symmetry operator from Eq.~\eqref{eq:ising-sym} and $\mathcal{P}_{\pm} = \frac{1}{2}(1\pm\mathcal{S})$ is the projector onto $\mathcal{H}_{\pm}$. Since $\mathcal{S}$ flips all $L$ spins on the chain, 
we can see from Eq.~\eqref{eq:trace-diff} that 
$\text{Tr}_{+}\{\mathcal{O}\} - \text{Tr}_{-}\{\mathcal{O}\}$ will vanish unless 
$\mathcal{O}$ acts nontrivially on all $L$ spins. 

We can now apply this reasoning to a typical term
\beq
	\frac{1}{Z_0}\text{Tr}_{+}\{e^{-\beta H_0}V_{Y_n}(\tau_n)\cdots V_{Y_1}(\tau_1)\}-\frac{1}{Z_0}\text{Tr}_{-}\{e^{-\beta H_0}V_{Y_n}(\tau_n)\cdots V_{Y_1}(\tau_1)\}
\eeq
that would appear in the (summand and integrand of the) difference $W_{+}(X) - W_{-}(X)$. Since each operator $V_Y$ acts on at most two spins, we see from Eq.~\eqref{eq:trace-diff} that 
this difference must vanish if $n$ is less than $L/2$. Thus, the difference of weights $W_{+}(X) - W_{-}(X)$ only receives contributions from terms of order $\lambda^n$ with $n \geq L/2$. This is because $n$ here keeps track of the order of the Duhamel expansion, and in that expansion the order $n$ term is accompanied by a factor of $\lambda^n$.
Finally, since the polymer expansion expresses $\ln(Z_{\pm}/Z_0)$ as a sum over products of
the weights $W_{\pm}(\chi)$, the same result
holds for the difference of free energies $f(\lambda, \beta) = -\frac{1}{\beta}\ln\left(Z_{+}\right) + \frac{1}{\beta}\ln\left(Z_{-}\right)$.
This can be proven order by order in the polymer expansion (since it converges
absolutely). 
This completes the proof of Claim~\ref{claim2} and of the second step of our proof of Theorem~\ref{thm:stability}.

\section{Generalization to other models}
\label{sec:generalizations}

Although we have focused on the specific model Hamiltonian from Sec.~\ref{sec:model}, it is
possible to generalize our stability result to a much larger family of Hamiltonians. In particular, 
our result also holds for models (i) without translation invariance, (ii) in dimensions greater than one, and
(iii) with a much wider variety of interaction terms. In particular, in regards to point (iii) we will
show that our result is not limited to the case of two-body interactions. 
In this section we briefly discuss a more general family of models that our main result also applies to.

The family of models that we consider here all involve spin $1/2$ degrees of freedom
in $D\geq 1$ spatial dimensions and on a (hyper)cubic lattice of linear 
size $L$. Let $\sigma^{x,y,z}_{\mb{r}}$ be the spin operators (really, the Pauli matrices) 
acting on site $\mb{r}$ of the lattice, and let $\mathcal{S} = \prod_{\mb{r}}\sigma^x_{\mb{r}}$ be the 
Ising symmetry operator. 
Our starting point is again an unperturbed Hamiltonian $H_0$ of the Ising form,
\beq
	H_0 = \frac{\Delta}{2}\sum_{\lan\mb{r},\mb{r}'\ran}(1-\sigma^z_{\mb{r}}\sigma^z_{\mb{r}'})\ ,
\eeq
where $\lan\mb{r},\mb{r}'\ran$ denotes a sum over nearest neighbors (counting each pair only once).
As before, $H_0$ commutes with $\mathcal{S}$ and has a unique ground state and a finite energy gap in 
each sector $\mathcal{H}_{\pm}$ of fixed $\mathcal{S}$ eigenvalue (we still have $\Delta > 0$). The
ground state $|\pm\ran$ in the sector $\mathcal{H}_{\pm}$ is again given by the appropriate
superposition of the all-up and all-down states, $|\pm\ran = (|\Uparrow\ran\pm|\Downarrow\ran)/\sqrt{2}$.

Our full Hamiltonian takes the form $H = H_0 + \lambda V$, where we assume that 
the perturbation $V$ preserves the Ising symmetry, $[\mathcal{S}, V] = 0$. 
The main difference from before is that we now allow $V$ to take the general form
\beq
	V = \sum_{\mb{r}} h_{\mb{r}} V_{\mb{r}}\ ,
\eeq
where the $h_{\mb{r}}$ are real coefficients satisfying $|h_{\mb{r}}| \leq h$ for some positive real
number $h$, and the $V_{\mb{r}}$ are operators that we now describe. 
Specifically, $V_{\mb{r}} = T_{\mb{r}}V_{\mb{0}}T_{\mb{r}}^{-1}$,
where $T_{\mb{r}}$ is the operator that performs translations by $\mb{r}$, and $V_{\mb{0}}$ is of the form
\beq
	V_{\mb{0}} = \sum_{Y} V_Y\ .
\eeq
where the sum runs over subsets $Y$ of lattice sites containing the origin $\mb{r}=\mb{0}$, and having at most $K$ sites, and where $V_Y$ is supported on $Y$. Note that in our setup we have broken translation symmetry because we allow each term $V_{\mb{r}}$ to appear with a different coefficient $h_{\mb{r}}$. Note also that $V_{\mb{0}}$ (and therefore $H$) contains at most $K$-body interactions. We assume that $K$ remains fixed and finite in the thermodynamic limit $L\to\infty$.

In this general case we need to impose two additional conditions on the interaction
terms $V_Y$ to guarantee that our results hold. The first condition is a generalization of the summability condition \eqref{eq:sum-bound}, namely 
\beq
	\sum_Y ||V_Y|| \leq c\ , \label{eq:general-sum-bound}
\eeq
where $||\cdot||$ is the usual operator norm and the constant $c$ is again required to be independent of
the lattice size. As before, this condition guarantees that the operator norm of the Hamiltonian
is extensive in the system size.

For the second condition, we require the $V_Y$ to obey an additional symmetry relation of the form
\beq
	[\mathcal{S}_{Y_i},V_Y] = 0 \ \ \forall\ i\ , \label{eq:symmetry-condition}
\eeq
where the $Y_i$ are the connected components of $Y$, and $\mathcal{S}_{Y_i}$ is the Ising symmetry operator
\emph{restricted to $Y_i$}: $\mathcal{S}_{Y_i} = \prod_{\mb{r}\in Y_i}\sigma^x_{\mb{r}}$.

To understand why we impose the condition \eqref{eq:symmetry-condition}, consider a simple $V$ that violates this condition, namely a long range $\sigma^z_{\mb{r}}\sigma^z_{\mb{r}'}$ interaction. It is known that (antiferromagnetic) interactions of this kind can cause a phase transition for arbitrarily small $\lambda$ even when the summability condition \eqref{eq:general-sum-bound} holds~\cite{daniels-vanEnter,vanEnter,biskup-chayes-kivelson}; therefore we cannot drop \eqref{eq:symmetry-condition} without imposing stricter requirements on the decay of the interactions. Furthermore, at a technical level, the condition \eqref{eq:symmetry-condition} is essential to our proof as it guarantees the factorization of the weights in our polymer model. We also note that \eqref{eq:symmetry-condition} is a natural property of many long-range interactions -- for example, those that are induced by integrating out gapless degrees of freedom (e.g., phonons or photons) that are even under the Ising symmetry.

For any $V$ of this kind, we can prove the following result:

\begin{thm}
\label{thm:generalized-stability}
There exists a $L$-independent constant $\lambda_0>0$
such that, if $|\lambda| < \lambda_0$, then (1) $H$
has a unique ground state and a finite energy gap in each sector $\mathcal{H}_{\pm}$ of the Hilbert
space with fixed $\mathcal{S}$ eigenvalue, and (2) the ground
state energy splitting $|E_{+}(\lambda)-E_{-}(\lambda)|$ between sectors satisfies the exponential bound
\beq
	|E_{+}(\lambda)-E_{-}(\lambda)| \leq c_1 L^D e^{-c_2 L^D}\ , \label{eq:exponential-bound-general}
\eeq
where $D\geq 1$ is the spatial dimension and $c_1,c_2$ are positive constants that depend on $\Delta$,
$\lambda$, $h$, and the form of the interactions, but not on the system size $L$. 
\end{thm}

In Appendix~\ref{app:generalizations} we summarize the changes that we need to make in our setup and
in the proofs of Lemma 1 and 2 in order to prove this more general result.

\section{Conclusion}
\label{sec:conclusion}

In this paper we investigated the stability of certain gapped Hamiltonians $H_0$ to perturbations
$V$ that contain long-range interactions, such as those that decay as a power law of the
distance. For the family of models that we considered we 
were able to prove not only the stability of the spectral gap, but also a more detailed stability property 
for the states below the gap. Specifically, for an $H_0$ with two exactly degenerate ground states below
the gap, we were able to prove that the residual splitting of these two states in the perturbed model was 
\emph{exponentially small} in the system size. As we mentioned in the introduction, this exponential
splitting bound is of great interest for quantum computation applications.

There are many possible directions for future work. One of the most interesting directions would
be to widen the class of unperturbed Hamiltonians $H_0$ that we can study. In this
paper we were mostly limited to the case where $H_0$ is classical (although the Jordan-Wigner transformation allowed us to prove results about some one dimensional quantum models, such as Kitaev's p-wave wire model~\cite{kitaev}). It would be interesting to consider the case where $H_0$ is a truly quantum Hamiltonian in a dimension higher than one, for example the toric code model in two dimensions. 

It would also be interesting to understand how to establish stability (if it does exist) in the
presence of long-range interactions that violate our summability condition, for example the 
$1/r$ Coulomb interaction in one or higher dimensions. To investigate that case it may be necessary to include the contributions to the Hamiltonian from the background charges that make the entire system neutral, as
is known to be important for proving the stability of matter~\cite{lieb-stability}.

\acknowledgments

We thank I. Klich and S. Michalakis for helpful conversations and for explaining their work to us.
We would like to thank A.C.D. van Enter for pointing out an instability associated with long-range 
antiferromagnetic $\sigma^z_{\mb{r}} \sigma^z_{\mb{r}'}$ interactions, which alerted us to an error in a 
previous statement of Theorem 3.
M.F.L. and M.L. acknowledge the support of the Kadanoff Center for Theoretical Physics at the University of 
Chicago. This work was supported by the Simons Collaboration on Ultra-Quantum Matter, which is a grant from the 
Simons Foundation (651440, M.L.).

\appendix

\section{Properties of the weights $W_{\pm}(X)$ and proof of Lemma 1}
\label{app:weights}

In this appendix we present a detailed analysis of the weights $W_{\pm}(X)$ for our model. We
first give a precise definition of $W_{\pm}(X)$. This definition can be thought of as
a more careful form of Eq.~\eqref{eq:sloppy-weight-def}.  
In the second part of this appendix
we then present the proof of Lemma~\ref{lem:weights}, which establishes the bound 
\eqref{eq:weight-bound} on $|W_{\pm}(X)|$. Finally, in the third part of this appendix we discuss two
small technical details that arise in the derivation of the bound on $|W_{\pm}(X)|$.

\subsection{Precise definition of $W_{\pm}(X)$}

To write down a precise expression for the weights we first 
decompose the weight $W_{\pm}(X)$ for any non-empty support set $X$ 
into a sum over contributions from all orders $n\geq 1$ in the Duhamel expansion of 
$Z_{\pm}$, and a sum over contributions from all possible subsets $Y_1,\dots,Y_n$ that determine 
the perturbation terms that appear at order $n$,
\beq
	W_{\pm}(X)= \sum_{n\geq 1}\sum_{Y_1,\dots,Y_n}W^{(n)}_{\pm;\ Y_1,\dots,Y_n}(X)\ . \label{eq:W-sum}
\eeq
With this decomposition, it is possible to write down a precise expression for the
weight $W^{(n)}_{\pm;\ Y_1,\dots,Y_n}(X)$ with the help of two \emph{indicator functions} that
we now define.

In order to compute $W^{(n)}_{\pm;\ Y_1,\dots,Y_n}(X)$, we
need to integrate over all times 
$0\leq \tau_1\leq \dots\leq\tau_n\leq \beta$ such that the pairs 
$(Y_1,\tau_1),\dots,(Y_n,\tau_n)$ are consistent with the
support set $X$. For fixed $Y_1,\dots,Y_n$, there may be several different ways of distributing the 
pairs $(Y_1,\tau_1),\dots,(Y_n,\tau_n)$ among the different time intervals of the blocked
spacetime lattice such that the resulting configuration is consistent with $X$. We will need
to sum/integrate over all of these consistent ways of distributing the pairs 
$(Y_1,\tau_1),\dots,(Y_n,\tau_n)$ in our final expression for $W^{(n)}_{\pm;\ Y_1,\dots,Y_n}(X)$.

Consider first the set of times where one or more of $\tau_1,\dots,\tau_n$
is an exact integer multiple of the temporal lattice spacing $\tau$. This set of times is a subset of $\mathbb{R}^n$ of Lebesgue measure zero, and so it does not actually contribute to the weight $W^{(n)}_{\pm;\ Y_1,\dots,Y_n}(X)$ (which involves an integral over $\tau_1,\dots,\tau_n$). Therefore, in studying
the distribution of the pairs $(Y_1,\tau_1),\dots,(Y_n,\tau_n)$ among the different time intervals,
we can restrict our attention to the cases where each of the times $\tau_1,\dots,\tau_n$ lies in an open 
interval $(\tau(\ell-1),\tau\ell)$ for some $\ell\in\{1,\dots,M\}$. We now proceed to consider these
cases.

A particular distribution of the $n$ pairs $(Y_1,\tau_1),\dots,(Y_n,\tau_n)$ among the open
time intervals $(\tau(\ell-1),\tau\ell)$ is described
by a set of non-negative integers $p_1,\dots,p_M$ satisfying 
$\sum_{\ell=1}^M p_{\ell}=n$ and such that the first 
$p_1$ pairs are located in the first (open) time slice $(0,\tau)$, the next $p_2$ pairs are located in the 
second time slice $(\tau,2\tau)$, and so on. 
Let $\vec{Y}= (Y_1,\dots,Y_n)$ denote the 
$n$-tuple of subsets, and let $\vec{p}=(p_1,\dots,p_M)$ denote the $M$-tuple of integers that determine
a particular distribution of the pairs $(Y_1,\tau_1),\dots,(Y_n,\tau_n)$ among the different time intervals.
In addition, let $\vec{\tau}=(\tau_1,\dots,\tau_n)$ be the $n$-tuple containing the times where
a perturbation acts in the integrand of a term at $n$th order in the Duhamel expansion.

To keep track of the distributions that are consistent with $X$, we define two indicator functions 
$I_X[\vec{Y};\vec{p}]$ and $I_{\vec{p}}[\vec{\tau}]$ as follows. 
The first function $I_X[\vec{Y};\vec{p}]$ is equal to $1$ if the set $\vec{Y}$ and the distribution 
described by $\vec{p}=(p_1,\dots,p_M)$ is consistent with $X$, and equal to $0$ otherwise. 
The second function $I_{\vec{p}}[\vec{\tau}]$ is equal to $1$ if the first $p_1$ times
in $\vec{\tau}$ lie in $(0,\tau)$, 
the next $p_2$ times in $\vec{\tau}$ lie in $(\tau,2\tau)$, etc., and equal to $0$ otherwise. 
With these notations we have
\begin{widetext}
\beq
	W^{(n)}_{\pm;\ Y_1,\dots,Y_n}(X)= \frac{(-\lambda)^n}{Z_{0}} \sum_{\substack{p_1,\dots,p_M\geq 0\\ p_1+\cdots+p_M=n}}I_X[\vec{Y};\vec{p}]\int_0^{\beta} d\tau_n\ \cdots\int_0^{\tau_2}d\tau_1\ I_{\vec{p}}[\vec{\tau}]\ \text{Tr}_{\pm}\left\{e^{-\beta H_0}V_{Y_n}(\tau_n)\cdots V_{Y_1}(\tau_1)\right\}\ , 
	\label{eq:precise-weights}
\eeq
\end{widetext}
where we sum over all allowed distributions $p_1,\dots,p_M$ of the pairs 
$(Y_1,\tau_1),\dots,(Y_n,\tau_n)$ into
the different time intervals. This expression for $W^{(n)}_{\pm;\ Y_1,\dots,Y_n}(X)$, combined
with Eq.~\eqref{eq:W-sum}, gives a precise definition of the weight $W_{\pm}(X)$ for each 
non-empty support set $X$. Note also that the empty support set $X=\varnothing$ can be
consistently assigned a weight of $1$, $W_{\pm}(\varnothing)=1$, and this particular contribution to 
$Z_{\pm}/Z_{0}$ comes from the zeroth order term in the Duhamel expansion of $Z_{\pm}$.

\subsection{Proof of Lemma 1}

We now prove the bound \eqref{eq:weight-bound} on $|W_{\pm}(X)|$ that is stated
in Lemma~\ref{lem:weights} from the main text. 
To prove this bound we start with the precise expression for the weights
$W^{(n)}_{\pm;\ Y_1,\dots,Y_n}(X)$ from \eqref{eq:precise-weights} above. 
To derive an upper bound on $|W^{(n)}_{\pm;\ Y_1,\dots,Y_n}(X)|$, we first use
the trivial bound $Z_{0}\geq 1$ to find that $(Z_{0})^{-1}\leq 1$ (and this bound is expected
to be reasonably tight in the low temperature regime that we are interested in). Next, we note that
\beq
	\text{Tr}_{\pm}\left\{e^{-\beta H_0}V_{Y_n}(\tau_n)\cdots V_{Y_1}(\tau_1)\right\} = \sum_{s_1,\dots,s_n} \lan s_n,\pm |V_{Y_n}|s_{n-1},\pm\ran\cdots\lan s_1,\pm|V_{Y_1}|s_n,\pm\ran  e^{-(\beta-\tau_n+\tau_1)\ep_{s_n}}\cdots e^{-(\tau_2-\tau_1)\ep_{s_1}}\ ,
\eeq
where $|s,\pm\ran$ are the eigenstates of $H_0$ that we introduced in the main text. 
Using the expression \eqref{eq:energies} for the energies $\ep_s$, we find that
\beqa
	e^{-(\beta-\tau_n+\tau_1)\ep_{s_n}}\cdots e^{-(\tau_2-\tau_1)\ep_{s_1}} &=& e^{-\Delta\times(\text{total vertical length of worldlines in } \mathcal{C})} \nnb \\
	&\leq& e^{-\Delta\tau p(X)}\ ,
\eeqa
where $\mathcal{C}$ is the microscopic configuration of 
worldlines determined by $(Y_1,\tau_1),\dots,(Y_n,\tau_n)$ and $|s_n,\pm\ran$, 
and $p(X)$ is the number of plaquettes in $X$. 
In addition, because of the time-ordering and the
grouping of the pairs $(Y_1,\tau_1),\dots,(Y_n,\tau_n)$, we have
\beq
	\int_0^{\beta} d\tau_n\ \cdots\int_0^{\tau_2}d\tau_1\ I_{\vec{p}}[\vec{\tau}]=\frac{\tau^n}{p_1!\cdots p_M!}\ .
\eeq

To proceed, let $n_1$ be the number of transverse field terms that appear
in the set $V_{Y_1},\dots,V_{Y_n}$ of perturbation terms. Similarly, let $n_{2,r}$
be the number of long-range perturbation terms acting at a distance $r$ that appear in the set
$V_{Y_1},\dots,V_{Y_n}$. [The subscripts 1 and 2 indicate that these keep track of one-body and two-body
terms, respectively.] Since there are $n$ perturbation terms in total, we have the identity
\beq
	n = n_1 + n_{2}\ , \quad \quad n_{2} = \sum_{r=1}^{\frac{L}{2}}n_{2,r}\ .
\eeq
If we use these definitions, and also use Eqs.~\eqref{eq:matrix-elements} to bound the matrix
elements of the perturbation terms $V_Y$, then we find the bound (note the plus subscript on the
trace on the right-hand side)
\begin{widetext}
\beq
	|W^{(n)}_{\pm; Y_1,\dots,Y_n}(X)|\leq (|\lambda| |h|\tau)^{n_1}(|\lambda|\tau)^{n_2}\left[\prod_{r=1}^{\frac{L}{2}} |f(r)|^{n_{2,r}} \right] e^{-\Delta\tau p(X)} \sum_{\substack{p_1,\dots,p_M\geq 0\\ p_1+\cdots+p_M=n}}I_X[\vec{Y};\vec{p}]\frac{1}{p_1!\cdots p_M!}\text{Tr}_{+}\left\{\td{V}_{Y_n}\cdots \td{V}_{Y_1}\right\}\ , 
	\label{eq:trace-plus-bound}
\eeq
\end{widetext}
where we also defined the operators $\td{V}_Y$ by
\beq
	\td{V}_Y = \begin{cases} \sigma^x_j\ & \text{ if }\ \ Y=\{j\}\ , \\ 
\sigma^x_j\sigma^x_k\ & \text{ if }\ \ Y=\{j,k\}\ .
\end{cases}
\eeq

We now simplify the prefactors that depend on $\lambda$ and $h$ in our bound on 
$|W^{(n)}_{\pm; Y_1,\dots,Y_n}(X)|$. We start by writing
\beqa
	(|\lambda|\tau)^{n_2} 
	&=& (\sqrt[4]{|\lambda|\tau})^{2 n_2}(\sqrt{|\lambda|\tau})^{n_2}\ .
\eeqa
We then have
\begin{align}
	(|\lambda||h|\tau)^{n_1}(|\lambda|\tau)^{n_2}\left[\prod_{r=1}^{\frac{L}{2}}|f(r)|^{n_{2,r}} \right] = (|\lambda||h|\tau)^{n_1}(\sqrt[4]{|\lambda|\tau})^{2 n_2}(\sqrt{|\lambda|\tau})^{n_2}
	\left[\prod_{r=1}^{\frac{L}{2}}|f(r)|^{n_{2,r}} \right]\ . 
\end{align}
Next, we assume that there exists some real number $\al_0 >0$ such that
\begin{align}
	|\lambda||h|\tau \leq \al_0, \quad \quad \sqrt[4]{|\lambda|\tau} \leq \al_0\ .
\end{align}
We will choose a particular value for $\al_0$ later. With this assumption we find that
\beq
	(|\lambda||h|\tau)^{n_1}(|\lambda|\tau)^{n_2}\left[\prod_{r=1}^{\frac{L}{2}}|f(r)|^{n_{2,r}} \right]  \leq (\al_0)^{n_1+2n_2}(\sqrt{|\lambda|\tau})^{n_2}\left[\prod_{r=1}^{\frac{L}{2}}|f(r)|^{n_{2,r}} \right]\ .
\eeq
If we now assume that $\al_0\leq 1$ and that $\sqrt{|\lambda|\tau}\leq 1$, and use
our assumption that $|f(r)|\leq 1$ for all $r$, then we have
\beq
	(|\lambda||h|\tau)^{n_1}(|\lambda|\tau)^{n_2}\left[\prod_{r=1}^{\frac{L}{2}}|f(r)|^{n_{2,r}} \right]  \leq (\al_0)^{b(X)}(\sqrt{|\lambda|\tau})^{d(X)}\left[\prod_{r=1}^{\frac{L}{2}}|f(r)|^{d_r(X)} \right]\ ,
\eeq
where we used the bounds
\begin{align}
	b(X) \leq n_1 + 2n_2, \quad \quad d_r(X) \leq n_{2,r}\ .
\end{align}
[Recall that $b(X)$ is the number of boxes in $X$ and $d_r(X)$ is the number of dashed lines of length
$r$ in $X$.]
At this point our bound on the weights takes the form
\begin{align}
	|W^{(n)}_{\pm;\ Y_1,\dots,Y_n}(X)|\leq (\al_0)^{b(X)}(\sqrt{|\lambda|\tau})^{d(X)}\left[\prod_{r=1}^{\frac{L}{2}}|f(r)|^{d_r(X)} \right]e^{-\Delta\tau p(X)}  \sum_{\substack{p_1,\dots,p_M\geq 0\\ p_1+\cdots+p_M=n}}I_X[\vec{Y};\vec{p}]\frac{1}{p_1!\cdots p_M!}\text{Tr}_{+}\left\{\td{V}_{Y_n}\cdots \td{V}_{Y_1}\right\}\ .
\end{align}

The last step in obtaining a bound on $|W_{\pm}(X)|$ is to sum over all choices of $Y_1,\dots,Y_n$, and
then to sum over all $n\geq 1$ (we sum over $n\geq 1$ because we are assuming
that $X$ is non-empty). In particular, we need a bound on 
\beq
	\sum_{n\geq 1} \sum_{Y_1,\dots,Y_n}\sum_{\substack{p_1,\dots,p_M\geq 0\\ p_1+\cdots+p_M=n}}I_X[\vec{Y};\vec{p}]\frac{1}{p_1!\cdots p_M!}\text{Tr}_{+}\left\{\td{V}_{Y_n}\cdots \td{V}_{Y_1}\right\}\ . 
	\label{eq:sum-to-bound}
\eeq
To obtain this bound we first define, for each time slice $(\tau(\ell-1),\tau\ell)$, a set 
$\mathcal{S}_{\ell}(X)$ of subsets $Y$ of lattice sites as follows. First, $\mathcal{S}_{\ell}(X)$
contains $\{j\}$ if the box centered at site $j$ in time slice $\ell$ is in $X$. Next,
$\mathcal{S}_{\ell}(X)$ contains $\{j,k\}$ if $X$ contains a dashed line between the boxes 
centered on sites $j$ and $k$ in time slice $\ell$. Using the sets $\mathcal{S}_{\ell}(X)$, 
we now define the ``partial perturbation terms'' $\mathcal{V}_{\ell}$ as
\beq
	\mathcal{V}_{\ell}= \sum_{Y\in \mathcal{S}_{\ell}(X)}\td{V}_Y\ .
\eeq

The next ingredient we need is a certain projector associated with the support set $X$ at
time zero. Let $X_0$ be the restriction of $X$ to the spatial slice at time zero.
By the construction of $X$, any bond $(j,j+1)$ at time zero that is contained in the \emph{complement} of $X_0$
must be in its ground state (i.e., no domain wall) for all microscopic configurations $\mathcal{C}$ with
the support set $X$. We define $\Pi_0$ to be the projector that projects these bonds onto their
ground state,
\beq
	\Pi_0= \prod_{(j,j+1)\subset X_0^c}\left(\frac{1+\sigma^z_j\sigma^z_{j+1}}{2}\right)\ .
\eeq

In terms of the projector $\Pi_0$ and the partial perturbation terms $\mathcal{V}_{\ell}$, the
sum from Eq.~\eqref{eq:sum-to-bound} can be bounded as
\begin{widetext}
\beq
	\sum_{n\geq 1} \sum_{Y_1,\dots,Y_n}\sum_{\substack{p_1,\dots,p_M\geq 0\\ p_1+\cdots+p_M=n}}I_X[\vec{Y};\vec{p}]\frac{1}{p_1!\cdots p_M!}\text{Tr}_{+}\left\{\td{V}_{Y_n}\cdots \td{V}_{Y_1}\right\} \leq \text{Tr}_{+}\left\{\Pi_0 e^{\mathcal{V}_1}e^{\mathcal{V}_2}\cdots e^{\mathcal{V}_M}\right\}\ ,
\eeq
\end{widetext}
and this bound is similar to the bound in Eq.~4.19 of KT~\cite{KT}. The way to understand this
bound is to note that the trace on the
right-hand side contains all terms on the left-hand side, and possibly even more terms. However,
any extra terms are still positive, so this is an upper bound. In addition, 
the factorials that appear on the
left-hand side are recovered after Taylor-expanding the exponentials $e^{\mathcal{V}_{\ell}}$ 
on the right-hand side.

The final step is to bound 
$\text{Tr}_{+}\left\{\Pi_0 e^{\mathcal{V}_1}e^{\mathcal{V}_2}\cdots e^{\mathcal{V}_M}\right\}$, 
and we can do this using the inequality from Eq.~\eqref{eq:plus-trace-bound}. 
To apply that inequality we choose $A=\Pi_0$ and 
$B= e^{\mathcal{V}_1}e^{\mathcal{V}_2}\cdots e^{\mathcal{V}_M}$. For $B$ we find that
\beq
	||B||\leq e^{\sum_{\ell=1}^M ||\mathcal{V}_{\ell}||} \leq e^{b(X)+d(X)}\ .
	\label{Bbound}
\eeq
Next, for $A$ we have
\beq
	\text{Tr}_{+}\{A\} \leq 2^{\text{\# of bonds overlapping with }X_0}\ .
\eeq
To simplify this factor, let $p_1(X)$ and $b_1(X)$ be the number of plaquettes and boxes, respectively, in 
$X$ in the first time slice. Then we have
\begin{align}
	\text{\# of bonds overlapping with } X_0 \leq p_1(X)+2 b_1(X)\ ,
	\label{bondbound}
\end{align}
which follows from the fact that each box overlaps with two adjacent bonds.

At this point our full bound on the weights takes the form
\begin{widetext}
\beqa
	|W_{\pm}(X)|&\leq& (\al_0)^{b(X)}(\sqrt{|\lambda|\tau})^{d(X)}\left[\prod_{r=1}^{\frac{L}{2}}|f(r)|^{d_r(X)} \right] e^{-\Delta\tau p(X)} e^{b(X)+d(X)} 2^{p_1(X)+2 b_1(X)} \nnb \\
	&\leq&  (\al_0)^{b(X)}e^{-\Delta\tau p(X)} e^{K|X|}(e\sqrt{|\lambda|\tau})^{d(X)}\left[\prod_{r=1}^{\frac{L}{2}}|f(r)|^{d_r(X)} \right]  \ ,
\eeqa
\end{widetext}
where $K$ is a numerical constant of order $1$.\footnote{For example, this bound will hold if we choose
$K=1+2\ln(2)$.}
Next, we make our choice for $\al_0$. We follow KT~\cite{KT} and choose
\beq
	\al_0= e^{-\Delta\tau}\ ,
\eeq
and so we find that
\beq
	|W_{\pm}(X)| \leq e^{-\Delta\tau |X| + K|X|}(e\sqrt{|\lambda|\tau})^{d(X)}\left[\prod_{r=1}^{\frac{L}{2}}|f(r)|^{d_r(X)} \right] \ .
\eeq
Now suppose that we are given some $\mu, \delta >0$, and we want to be able to satisfy the bound
from \eqref{eq:weight-bound}.
If we first choose $\tau$ large enough to satisfy
\beq
	\tau\geq \frac{\mu+K}{\Delta}\ , \label{eq:tau-condition}
\eeq
and then choose $\lambda$ small enough to satisfy 
\beq
	e\sqrt{|\lambda|\tau}\leq \delta\ ,
\eeq
then we find the desired bound
\beq
	|W_{\pm}(X)| \leq e^{-\mu |X|}\delta^{d(X)}\left[\prod_{r=1}^{\frac{L}{2}}|f(r)|^{d_r(X)} \right] \ .
\eeq

Finally, from our assumptions that $|\lambda||h|\tau \leq \al_0$ and $\sqrt[4]{|\lambda|\tau} \leq \al_0$,
we find that our bound on $|W_{\pm}(X)|$ holds for all $\lambda\in\mathbb{C}$ 
satisfying $|\lambda|\leq \lambda_0$, where 
\beq
	\lambda_0 = \min\left(\frac{1}{|h|\tau}e^{-\Delta\tau},\ \frac{1}{\tau}e^{-4\Delta\tau},\ \frac{1}{\tau}\frac{\delta^2}{e^2}\right)\ ,
\eeq
and the value of $\tau$ is determined by \eqref{eq:tau-condition}.
It is crucial for our results that $\lambda_0$ is independent of 
the system size $L$ and the inverse temperature $\beta$. 
Note also that $\al_0\leq 1$ by construction, since $\Delta,\tau\geq 0$. 
In addition, if $|\lambda|\leq \lambda_0$, then we also have $\sqrt{|\lambda|\tau}\leq 1$.
This completes the proof of Lemma~\ref{lem:weights}.

\subsection{Technical details for the bound on the weights}

We close this appendix by discussing two small technical details that arose in our 
derivation of the bound on $W_{\pm}(X)$. 

\subsubsection{Signs of matrix elements in Ising symmetry sectors}

The first technical detail that we address here is the signs of the matrix elements
$\lan s,\pm |V_Y|s',\pm\ran$ of the perturbation terms $V_Y$, where $|s,\pm\ran$ and $|s',\pm\ran$ are
two of the eigenstates of $H_0$ in the sector $\mathcal{H}_{\pm}$, defined in Sec.~\ref{sec:setup}. 

Since $V_Y$ contains either a single operator $\sigma^x_j$ or a product $\sigma^x_j\sigma^x_k$ of two
Pauli $x$ operators, we need to consider the signs of matrix elements of $\sigma^x_j$ and
$\sigma^x_j\sigma^x_k$ between two states $|s,\pm\ran$ and $|s',\pm\ran$. Using
$[\mathcal{S},\sigma^x_j]=0$ and $\mathcal{S}^2=1$, we find that
\beqa
	\lan s,\pm|\sigma^x_j |s',\pm\ran &=& \lan s|\sigma^x_j |s'\ran \pm \lan s|\sigma^x_j \mathcal{S}|s'\ran \\
	\lan s,\pm|\sigma^x_j\sigma^x_k|s',\pm\ran &=& \lan s|\sigma^x_j\sigma^x_k |s'\ran \pm \lan s|\sigma^x_j\sigma^x_k \mathcal{S}|s'\ran \ .
\eeqa
The key point here is that, for given $|s\ran$ and $|s'\ran$, 
only one of the terms on the right-hand sides of 
these equations can be non-zero. Therefore we find that
\begin{subequations}
\label{eq:matrix-elements}
\beqa
	|\lan s,-|\sigma^x_j |s',-\ran| &=& \lan s,+|\sigma^x_j |s',+\ran \\
	|\lan s,-|\sigma^x_j\sigma^x_k |s',-\ran| &=& \lan s,+|\sigma^x_j\sigma^x_k |s',+\ran\ .
\eeqa
\end{subequations}
[The matrix elements of $\sigma^x_j$ and $\sigma^x_j\sigma^x_k$ in ``$+$'' states are always positive.]
These relations are the reason why there is only a ``$+$'' subscript on the trace in the bound in
Eq.~\eqref{eq:trace-plus-bound}.

\subsubsection{Modified trace bound}

The next issue that we address here is the problem of obtaining upper bounds on traces
of the form $\text{Tr}_{+}\{AB\}$, where the
trace is taken over the sector $\mathcal{H}_{+}$ of the total Hilbert space. 
In our derivation of the bound on $|W_{\pm}(X)|$, we encountered a trace of this
form where both operators $A$ and $B$ commuted with the Ising symmetry operator $\mathcal{S}$, and where
the operator $A$ was semipositive definite. In that case it is possible
to obtain a useful bound on $\text{Tr}_{+}\{AB\}$ in the following way. Let $|n,+\ran$
be a basis of simultaneous eigenstates of $\mathcal{S}$ and $A$ in the plus sector, with
$\mathcal{S}|n,+\ran=|n,+\ran$ and $A|n,+\ran=a_n|n,+\ran$, where $a_n\geq 0$ are the
eigenvalues of $A$ within $\mathcal{H}_{+}$. Then we have
\beqa
	\text{Tr}_{+}\{AB\} = \sum_{n}a_n \lan n,+|B|n,+\ran \nnb \leq ||B ||\sum_{n}a_n \ .
\eeqa
Since $\sum_n a_n= \text{Tr}_{+}\{A\}$, we find that $\text{Tr}_{+}\{AB\}$ is bounded
as
\beq
	\text{Tr}_{+}\{AB\} \leq ||B || \text{Tr}_{+}\{A\} \ . \label{eq:plus-trace-bound}
\eeq
We used this inequality to bound 
$\text{Tr}_{+}\left\{\Pi_0 e^{\mathcal{V}_1}e^{\mathcal{V}_2}\cdots e^{\mathcal{V}_M}\right\}$ in Eqs.~(\ref{Bbound})-(\ref{bondbound}).

\section{Proof of Lemma 2}
\label{app:combinatorics}

Our goal in this appendix is to prove Lemma~\ref{lem:comb}, which establishes sufficient conditions
for the bound $q < 1$ to hold, where
\beq
	q = \sum_{\gamma\ni v}\td{W}(\gamma)e^{14|\gamma|} \ . \label{Z}
\eeq
The sum is taken over all polymers that contain the particular vertex $v$, and
the weight $\td{W}(\gamma)$ is given by
\beq
	\td{W}(\gamma)= e^{-\mu |\gamma|}\delta^{d(\gamma)}\left[\prod_{r=1}^{\frac{L}{2}}|f(r)|^{d_r(\gamma)} \right]\ . 
\eeq
This lemma is the key to proving the convergence of the polymer expansion for our model.

To gain some intuition for why it is still possible to prove that 
$q < 1$ with long-range interactions, and for why there must be some restriction on the form of $f(r)$, 
it is useful to look at a simple example. 
Suppose we have a weakly-connected polymer $\gamma$ which is made up of two connected polymers 
$\td{\gamma}_1$ and 
$\td{\gamma}_2$ that are connected to each other by a single dashed line, and where $v\in\td{\gamma}_1$. 
The contribution of this polymer to $q$ involves (among other things) a sum over the length $r$ of the single
dashed line that connects $\td{\gamma}_1$ and $\td{\gamma}_2$, with each term in the sum weighted by a factor of 
$|f(r)|$. If $f(r)$ satisfies the
summability condition \eqref{eq:sum-bound}, then this contribution is bounded by a finite constant (even
as $L\to\infty$), and we might expect that more complicated contributions to $q$ will also be finite. On
the other hand, if $\sum_r |f(r)|$ diverges as $L\to\infty$, then this simple example already shows that
we can never have $q<1$, and so the polymer expansion will not converge and the system
is likely to be unstable. 

We now move on to the formal proof of Lemma~\ref{lem:comb}.

\subsubsection{Setting up the calculation}

We start by introducing some notation. For each polymer $\gamma$, 
we define $n_c(\gamma)$ to be the number of 
connected components in $\Lambda_{\gamma}$ (the induced subgraph of $\Lambda$ determined by the vertices in 
$\gamma$).
It is convenient to organize the sum in (\ref{Z}) in terms of the number of connected components, 
$N \equiv n_c(\gamma)$, and the total size of the polymer, 
$\ell \equiv |\gamma|$. Defining
\begin{align}
	q(N,\ell) = \sum_{\substack{\gamma\ni v\\ n_c(\gamma)=N \\ |\gamma|=\ell}}\td{W}(\gamma)e^{14|\gamma|}
\end{align}
we have
\begin{align}
q = \sum_{\ell=1}^\infty \sum_{N=1}^{\ell} q(N,\ell)\ .
\end{align}
Here, the upper limit in the second sum is $\ell$ because a polymer of size $\ell$ can have at most 
$N \leq \ell$ connected components.

\subsubsection{Bounding $q(N,\ell)$ in terms of $q_{\mathrm{tree}}(N,\ell)$}

We now bound $q(N,\ell)$ in terms of a closely related quantity $q_{\text{tree}}(N,\ell)$. To define the 
latter quantity, note that there is a natural way to define a \emph{graph}\footnote{This graph is yet another
graph that should not be confused with the graphs that appear in the polymer expansion of
$\ln(Z)$ or the graph structure that we have added to $\Lambda$.} 
associated with each polymer $\gamma$.
The nodes of this graph correspond to the 
connected components of $\gamma$ after we erase the dashed lines (i.e., the connected
components of $\Lambda_{\gamma}$), while the edges correspond 
to the dashed lines. We will say that a polymer $\gamma$ is \emph{tree-like} if the 
corresponding graph is a tree: i.e. there is exactly 
one way to get from any connected component of $\gamma$ to any other connected
component of $\gamma$ by traversing dashed lines. 

We define $q_{\text{tree}}(N,\ell)$ to be a sum over tree-like $\gamma$ with $N$ connected components and 
total size $\ell$:
\begin{align}
	q_{\text{tree}}(N,\ell) = \sum_{\substack{\gamma\ni v\\ \text{$\gamma$ tree-like}\\ n_c(\gamma)=N \\ |\gamma|=\ell}}\td{W}(\gamma)e^{14|\gamma|}\ .
\end{align}

To derive a bound on $q(N,\ell)$ in terms of $q_{\text{tree}}(N,\ell)$, note that, given any 
polymer $\gamma$, we can always obtain a tree-like configuration by removing an appropriate subset of the 
dashed lines. Turning this around, it follows that we can construct every polymer $\gamma$ by starting 
with an appropriate tree-like configuration and then \emph{adding} 
dashed lines. 

The key question is then: how many ways can we add dashed lines to a tree-like $\gamma$ of size $\ell$? 
To answer this question, notice that dashed lines in some $\gamma$ can be parameterized by two vertices 
$v, v' \in \gamma$ that share the same time-coordinate: 
$\tau(v) = \tau(v')$. (Here, $\tau(v)$ denotes the time coordinate of $v$). If we imagine choosing $v$ first 
and $v'$ second, then it is clear that there are at most $\ell$ possibilities for $v$, and at most $L$ 
possibilities for $v'$ (since it must have the same time coordinate as $v$, and in addition the 
dashed lines only connect boxes to boxes or plaquettes to plaquettes, but not boxes to plaquettes).
It follows that the maximum number of possible dashed lines is $\ell L$.
Then, since there are at most $\ell L$ possible dashed lines that can be added to $\gamma$, 
the total number of 
ways to decorate $\gamma$ by dashed lines is at most $2^{\ell L}$ since we can choose to add or not add 
each possible dashed line. 

The above counting argument suggests that $q(N,\ell) \leq 2^{\ell L} q_{\text{tree}}(N,\ell)$, but this bound 
does not use the fact that a dashed line of length $r$ comes with a factor of $\delta |f(r)|$ (according
the weight $\td{W}(\gamma)$). 
Let us again consider the number of ways we can add dashed lines to a tree-like $\gamma$ of
size $\ell$. There are again at most $\ell$ choices for the first end of the dashed line, and at most $L-1$ 
choices for the second end (not $L$ choices since the dashed line must end in a different place than 
it started). We again have the option to add or not add each dashed line. However, each dashed line 
comes with a different weight according to its length. Suppose we have chosen a fixed location for the
vertex at the first end of a dashed line. Then choosing to add (with
weight $\delta |f(r)|$) or not add (with weight $1$) each dashed line with this 
fixed starting vertex yields a factor of
\beq
	\left[\prod_{r=1}^{\frac{L}{2}-1}\left(1+\delta |f(r)|\right)\right]^2 \left[1+\delta |f(\tfrac{L}{2})|\right] \ ,
\eeq
where the first factor in this equation is squared because we can go either way around the spatial circle
to find the second end of the dashed line (and there is only one dashed line with the maximum length of
$L/2$). It is convenient to now bound this factor from above
as
\beq
	\left[\prod_{r=1}^{\frac{L}{2}-1}\left(1+\delta |f(r)|\right)\right]^2 \left[1+\delta |f(\tfrac{L}{2})|\right] \leq \left[\prod_{r=1}^{\frac{L}{2}}\left(1+\delta |f(r)|\right)\right]^2 \ .
\eeq
This is the factor that we obtain for a fixed location of the vertex at the first end of 
the dashed line. We already mentioned that there are $\ell$ choices for the location of this first
vertex, and so in the end we find a bound with this factor raised to the $\ell$th power,
\beq
	q(N,\ell) \leq \left[\prod_{r=1}^{\frac{L}{2}}\left(1+\delta |f(r)|\right)\right]^{2\ell}
	q_{\text{tree}}(N,\ell)\ .
\eeq
Finally, for later use we note that we can use the bound $1+x\leq e^x$ and the summability condition
\eqref{eq:sum-bound} for $f(r)$ to find that
\beq
	\left[\prod_{r=1}^{\frac{L}{2}}\left(1+\delta |f(r)|\right)\right]^{2\ell} \leq e^{2\ell\delta\sum_{r=1}^{\frac{L}{2}}|f(r)|} \leq e^{\ell\delta c}\ .
\eeq
This leads us to a final bound on $q(N,\ell)$ of the form
\begin{align}
q(N,\ell) \leq e^{\ell\delta c} q_{\text{tree}}(N,\ell)\ .
\label{ZZtree}
\end{align}

\subsubsection{Bounding $q_{\mathrm{tree}}(N,\ell)$}

Our task is now to bound $q_{\text{tree}}(N,\ell)$. To this end, we now describe a way to ``grow'' a 
tree-like polymer starting with some more basic components. To explain our construction, we first need to 
fix an \emph{ordering} on the vertices in our graph $\Lambda$. 
This ordering can be arbitrary -- i.e. it need not have any geometric significance.

Our ``growing'' procedure takes several pieces of data as input: 
\begin{enumerate}
\item{An $N$-tuple of connected polymers $(\td{\gamma}_1,...,\td{\gamma}_N)$, each of which contain the distinguished vertex $v$ and none of which have any dashed lines. This collection of polymers should have total size $\ell$: i.e. $\sum_{i=1}^N\ell_i = \ell$ where 
$\ell_i = |\td{\gamma}_i|$.} 
\item{An $(N-1)$-tuple of vertices $(\ov{s}_1,\dots,\ov{s}_{N-1})$, where
$\ov{s}_k$ (for $k\in\{1,\dots,N-1\}$) belongs to the connected polymer $\td{\gamma}_{i_k}$ for
some integer $i_k$ satisfying $1\leq i_k \leq k$. 
We also assume that the pairs $(\ov{s}_k,i_k)$ form an increasing 
sequence with respect to an ordering ``$\prec$'' defined by
$(\ov{s}_k,i_k) \prec (\ov{s}_{k'},i_{k'})$ if either (1) $i_k < i_{k'}$, or (2) $i_k=i_{k'}$ and 
$\ov{s}_{k'}$ has a larger number than 
$\ov{s}_k$ in the original ordering of the vertices on $\Lambda$.}
\item{An $(N-1)$-tuple $(x_2,...,x_{N})$ of spatial (i.e., horizontal) displacements.}
\end{enumerate}

Given this input data, we can construct a corresponding tree-like polymer $\gamma$ as follows. First, we 
define $(2N-1)$ vertices, $\{s_1,...,s_{N-1}, v_1, v_2,...,v_N\}$ by setting $v_1 = v$ and using the following recursive equations:
\begin{align}  
s_k &= \bar{s}_k + (v_{i_k} - v) \nonumber \\
v_{k+1} &= s_k + (x_{k+1},0)\ .
\end{align}
Here in the first equation, we define $s_k$ by translating $\bar{s}_k$ by the vector $v_{i_k} - v$. In the 
second equation, we define $v_{k+1}$ by translating $s_k$ in the spatial direction by the amount 
$x_{k+1}$ (the notation $(x_{k+1},0)$ means a vector in the two-dimensional spacetime 
with spatial component equal to $x_{k+1}$ and temporal component equal to $0$).

To build our configuration $\gamma$, we then take the union
\begin{align}
 \bigcup_{k=1}^N [\td{\gamma}_k + (v_k-v)]
\end{align}
where the above notation means that we translate each $\td{\gamma}_k$ by the vector $v_k -v$ before taking 
the union. We then add dashed lines between the pairs $(s_k, v_{k+1})$ for $k\in\{1,\dots,N-1\}$. Note that $s_k$ and 
$v_{k+1}$ have the same time-coordinate by construction. Also $v_k \in \td{\gamma}_k + (v_k-v)$ and 
$s_k \in \td{\gamma}_{i_k} + (v_{i_k}-v)$.

Note that this construction can generate illegal polymers in some cases because the different connected components, $[\td{\gamma}_k + (v_k-v)]$, may overlap with one another. However, this is not important for our purposes: the only property that we need is the converse result that every tree-like polymer with $N$ connected components can be generated by this procedure. This property implies that the number of tree-like polymers $\gamma$ containing the distinguished vertex $v$ is less than or equal to the number of possible choices of the above data.

We now use this property of our construction to obtain an upper-bound on 
$q_{\text{tree}}(N,\ell)$. To this end, we first define
\begin{align}
q_0(\ell) 
&=  \sum_{\substack{\td{\gamma} \ni v \\ |\td{\gamma}| = \ell}} e^{-\mu|\td{\gamma}|} e^{14|\td{\gamma}|}\,
\end{align}
where the sum is taken over all \emph{connected} polymers $\td{\gamma}$ of size $\ell$ containing
the vertex $v$ and not having any dashed lines. With this notation, our construction implies the following upper bound on $q_{\text{tree}}(N,\ell)$:
\beq
	q_{\text{tree}}(N,\ell)\leq \sum_{\substack{\ell_1,\dots,\ell_N \\ \ell_1+\dots+\ell_N=\ell}}\prod_{i=1}^N q_0(\ell_i) \sum_{\substack{\ov{s}_1,\dots,\ov{s}_{N-1} \\ (\ov{s}_k,i_k) \text{increasing} \\
	\text{w.r.t. ``$\prec$''}}}\sum_{x_2,\dots,x_N}\delta^{N-1}\prod_{k=2}^{N}|f(|x_k|_{\text{p}})|\ . \label{eq:growing-procedure-bound}
\eeq
We now explain the different factors in this expression. The first part 
$\sum_{\ell_1,\dots,\ell_N}\prod_{i=1}^N q_0(\ell_i)$
accounts for the sum over different connected components $\td{\gamma}_1,...,\td{\gamma}_N$ of lengths 
$\ell_1,..., \ell_N$, and with total length $\ell$. 
Next, we sum over all $(N-1)$-tuples $(\ov{s}_1,\dots,\ov{s}_{N-1})$ of vertices that satisfy 
the properties discussed 
in item 2 in the list of data required for our tree-growing procedure. 
Next, we 
sum over all the spatial displacements $(x_2,\dots,x_N)$. Finally, the factor of 
$\delta^{N-1}\prod_{k=2}^{N}|f(|x_k|_{\text{p}})|$ assigns the
correct weight to the $N-1$ dashed lines in the tree-like configurations that we are generating.

Using the summability condition \eqref{eq:sum-bound} for $f(r)$, we find that
$\sum_{x_k}|f(|x_k|_{\text{p}})| \leq c$ for all $k$, and so
\begin{align}
	\sum_{\substack{\ov{s}_1,\dots,\ov{s}_{N-1} \\ (\ov{s}_k,i_k) \text{increasing} \\
	\text{w.r.t. ``$\prec$''}}}\sum_{x_2,\dots,x_N}\delta^{N-1}\prod_{k=2}^{N}|f(|x_k|_{\text{p}})| \leq \sum_{\substack{\ov{s}_1,\dots,\ov{s}_{N-1} \\ (\ov{s}_k,i_k) \text{increasing} \\
	\text{w.r.t. ``$\prec$''}}} (\delta c)^{N-1} \leq \binom{\ell+N-2}{N-1}(\delta c)^{N-1}\ ,
\end{align}
where $\binom{\ell+N-2}{N-1}$ is the number of increasing (but not necessarily strictly increasing) 
sequences of length $N-1$ taken from a set of size $\ell$. 
Using these results, we obtain a bound on $q_{\text{tree}}(N,\ell)$ of the form
\begin{align}
	q_{\text{tree}}(N,\ell)\leq \sum_{\substack{\ell_1,\dots,\ell_N \\ \ell_1+\dots+\ell_N=\ell}}\prod_{i=1}^N q_0(\ell_i)\binom{\ell+N-2}{N-1}(\delta c)^{N-1} \ . \label{Ztree1}
\end{align}

To proceed further we note that the partition function $q_0(\ell)$ obeys a bound of the form
\begin{align}
q_0(\ell) \leq e^{(K'-\mu) \ell} \label{eq:connected-bound}
\end{align}
where $K'$ is some constant of order $1$. This bound follows from the fact that the number of connected polymers $\tilde{\gamma}$ of size $\ell$ that contain $v$ can be bounded by 
$e^{K'' \ell}$ for some $K''$. The latter bound is a standard combinatorial result about the counting of connected subsets of a graph; see e.g. Lemma V.8.5 of Ref.~\onlinecite{simon}, which gives $K'' = 2 \ln(14)$,
and then $K' = 2\ln(14) + 14$ for our polymer model.

Substituting this bound into (\ref{Ztree1}) and simplifying, we derive
\begin{align}
q_{\text{tree}}(N,\ell) \leq \sum_{\substack{\ell_1,\dots,\ell_N \\ \ell_1+\dots+\ell_N=\ell}} \bpm \ell +N-2\\ N-1 \epm  \cdot (\delta c)^{N-1} \cdot e^{(K'-\mu) \ell}
\end{align}
Next we note that the number of $N$-tuples $(\ell_1,...,\ell_N)$ of positive integers with $\ell_1 +... + \ell_N = \ell$ is given by $\bpm \ell-1 \\ N-1 \epm$. Therefore, evaluating the sum gives
\begin{align}
q_{\text{tree}}(N,\ell) \leq \bpm \ell-1 \\ N-1 \epm \cdot \bpm \ell +N-2\\ N-1 \epm  \cdot (\delta c)^{N-1} \cdot e^{(K'-\mu) \ell}\ .
\label{Ztree2}
\end{align}
To simplify this expression we use the two inequalities
\begin{align}
\bpm \ell-1 \\ N-1 \epm \leq 2^\ell, \quad \quad  \bpm \ell +N-2\\ N-1 \epm &\leq   \bpm 2\ell \\ N-1 \epm
\end{align}
where the second inequality follows from the fact that $N \leq \ell$. Substituting these inequalities into (\ref{Ztree2}) gives
\begin{align}
q_{\text{tree}}(N,\ell) \leq \bpm 2\ell \\ N-1 \epm (\delta c)^{N-1} \cdot e^{(K'-\mu + \ln(2)) \ell}
\label{Ztree3}
\end{align}

\subsubsection{Bounding  $q$}

We are now ready to bound $q$. First, we combine (\ref{ZZtree}) and (\ref{Ztree3}) to derive
\begin{align}
q(N,\ell) \leq \bpm 2\ell \\ N-1 \epm (\delta c)^{N-1} \cdot e^{(K'-\mu + \delta c + \ln(2)) \ell} \ .
\end{align}
Next, we use the binomial expansion to deduce
\begin{align}
\sum_{N=1}^\ell q(N,\ell) \leq (1+\delta c)^{2\ell}  \cdot e^{(K'-\mu + \delta c + \ln(2) ) \ell}\ .
\end{align}
Applying the inequality $(1+x)^n \leq e^{nx}$, we derive
\begin{align}
\sum_{N=1}^\ell q(N,\ell) \leq  e^{(K'-\mu + 3\delta c + \ln(2)) \ell}\ .
\end{align}
It follows that
\begin{align}
q \leq \sum_{\ell=1}^\infty e^{(K'-\mu + 3\delta c + \ln(2)) \ell} 
\end{align}
Clearly the right hand side is a geometric series so we can sum it exactly. In particular we can see that 
$q < 1$ if we choose $e^{(K'-\mu + 3\delta c + \ln(2))} < 1/2$. This corresponds to the requirement that
\beq
	\mu - 3\delta c > K_0\ , \quad \quad K_0 = K' + 2\ln(2)\ .
\eeq
This completes the proof of Lemma~\ref{lem:comb}.

\section{Convergence criteria for the polymer expansion}
\label{app:polymer}

In this appendix we present more details on the different convergence criteria for the polymer 
expansion. We first explain how the
simplified convergence criterion that we presented in 
Sec.~\ref{sec:polymer} follows from a general convergence criterion that is derived 
in Ref.~\onlinecite{brydges}. We then explain how to derive the modified convergence criterion
that we used in Sec.~\ref{sec:lemma2} to handle the case where the notion of intersection of polymers is 
more complicated than simple sharing of vertices of $\Lambda$.

\subsection{Derivation of the simplified convergence criterion}

As in Sec.~\ref{sec:polymer}, we consider the setting where each polymer contains a subset of the 
vertices in the set $\Lambda$, and where two polymers are defined to intersect if they have one or more 
vertices in common. In this case, it is possible to derive a general criterion for the convergence of the
polymer expansion~\cite{brydges}. To state this general criterion we first
define, for any integer $d\geq 1$, a positive real number $\td{Q}(d)$ by
\beq
	\td{Q}(d):= \sup_{v\in\Lambda} \sum_{\gamma\ni v}|W(\gamma)|\ |\gamma|^{d-1}\ , \label{eq:Q-tilde-def}
\eeq
where the sum is taken over all polymers $\gamma$ that contain the vertex $v$, and then 
we take the supremum over all vertices $v$ in $\Lambda$. In terms of the $\td{Q}(d)$, 
we also define the positive real number $\td{q}$ by
\beq
	\td{q} := \sum_{d=1}^{\infty}\frac{1}{(d-1)!}\td{Q}(d)\ . \label{eq:q-tilde-def}
\eeq
In terms of $\td{q}$, one can show that for all $k\geq 1$, the $k$th term $I_k$ in 
the expansion \eqref{eq:polymer-expansion} of $\ln(Z)$ satisfies the bound
\beq
	|I_k| \leq \frac{1}{k}\td{q}^k |\Lambda|\ .
\eeq
We can see from this expression that the polymer expansion for $\ln(Z)$ 
will be absolutely convergent if $\td{q}<1$ (again, for a fixed value of $|\Lambda|$).
Therefore, in this general case we have the following result.

\begin{thm*}[Theorem 3.4 of Ref.~\onlinecite{brydges}]
\label{thm:convergence-general}
If $\td{q} < 1$, then the polymer expansion \eqref{eq:polymer-expansion} for $\ln(Z)$ is absolutely 
convergent for fixed $|\Lambda|$ and furthermore $\ln(Z)$ satisfies the bound
\beq
	|\ln(Z)|\leq \sum_{k=1}^{\infty}|I_k| \leq -|\Lambda|\ln(1-\td{q})\ .
\eeq
\end{thm*}

Using this general criterion, we now derive the simplified convergence criterion stated in 
Theorem~\ref{thm:convergence}. 
Recall that the simplified criterion applies in the case where we have an 
upper bound on the weights of the form $|W(\gamma)| \leq \td{W}(\gamma)$, and
where the set of polymers is \emph{isotropic} with respect to $\td{W}(\gamma)$ in the sense that the sum $\sum_{\gamma\ni v}\td{W}(\gamma)g(|\gamma|)$
is independent of $v$ for any function $g(|\gamma|)$ of $|\gamma|$.
In this case we can see that
\beq
	\sum_{\gamma\ni v}|W(\gamma)|\ |\gamma|^{d-1} \leq \sum_{\gamma\ni v}\td{W}(\gamma)\ |\gamma|^{d-1}\ ,
\eeq
and the value of the sum on the right-hand side of this inequality no longer
depends on the specific vertex $v$. Then for all $d$ we have 
\beq
	\td{Q}(d) \leq \sum_{\gamma\ni v}\td{W}(\gamma)\ |\gamma|^{d-1}\ ,
\eeq
with no supremum on the right-hand side, and so $\td{q} \leq q$, where $q = \sum_{\gamma\ni v}\td{W}(\gamma)e^{|\gamma|}$ was defined in Eq.~\eqref{eq:q-def} of the main text. Since $\td{q} \leq q$, we can prove the convergence of the polymer expansion
if we can verify our simplified convergence criterion, i.e., if we can prove that
$q < 1$.
 
\subsection{Derivation of the modified convergence criterion}

We now explain the origin of the modified convergence criterion that we used in Sec.~\ref{sec:lemma2}.
Recall that we needed to use this modified criterion because in our model the notion of intersection of 
polymers is more complicated than simple sharing of vertices of $\Lambda$. 

In the derivation of the general convergence criterion for the polymer expansion, as discussed in 
Ref.~\onlinecite{brydges}, the specifics of the notion of intersection of polymers is only 
used near the end of the proof. Specifically, this information is used to bound sums of the 
form
\beq
	\sum_{\substack{\gamma'\\\gamma' \cap \gamma \neq \varnothing}}|W(\gamma')|g(|\gamma'|)\ , \label{eq:overlap-sum}
\eeq
where the sum is taken over all polymers $\gamma'$ that intersect with a particular fixed polymer 
$\gamma$, and $g(|\gamma'|) \geq 0$ is a positive function of the polymer size $|\gamma'|$.

Consider first the simple case where the notion of intersection is equivalent to sharing vertices. In this
case we can follow Ref.~\onlinecite{brydges} and bound the sum \eqref{eq:overlap-sum} as
\beq
		\sum_{\substack{\gamma'\\\gamma' \cap \gamma \neq \varnothing}}|W(\gamma')|g(|\gamma'|) \leq \sum_{v \in\gamma} \sum_{\gamma'\ni v}|W(\gamma')|g(|\gamma'|)\ ,
\eeq
where we sum over all vertices $v$ in $\gamma$, and then we have an inner sum over all $\gamma'$ that contain 
$v$. Now suppose that we have a bound $|W(\gamma)| \leq \td{W}(\gamma)$, 
and that $\Gamma$ is isotropic with respect to $\td{W}(\gamma)$. Then we can proceed further with this bound to obtain
\beqa
	\sum_{\substack{\gamma'\\\gamma' \cap \gamma \neq \varnothing}}|W(\gamma')|g(|\gamma'|) &\leq& \sum_{v \in\gamma} \sum_{\gamma'\ni v}\td{W}(\gamma') g(|\gamma'|) \nnb \\
	&=& |\gamma| \sum_{\gamma'\ni v}\td{W}(\gamma') g(|\gamma'|)\ ,
\eeqa
where the last equality holds because our additional assumptions imply that the inner sum is independent
of $v$. If we continue to follow the proof in Ref.~\onlinecite{brydges}, and use this method to 
bound overlap sums like \eqref{eq:overlap-sum}, then we will recover the simplified convergence criterion 
that we presented in Theorem~\ref{thm:convergence} of Sec.~\ref{sec:polymer}.

Next, we discuss the modification to this procedure that we need for our model, where two polymers
$\gamma_1$ and $\gamma_2$ intersect if (i) they have one or more vertices in common, or 
(ii) a vertex in $\gamma_1$ is connected to a vertex in
$\gamma_2$ via an edge in $\Lambda$. To handle this case we define, for each polymer $\gamma$, a 
\emph{thickened} polymer $\gamma_{\text{thick}}$, where $\gamma_{\text{thick}}$ satisfies:
\begin{enumerate}
\item $\gamma_{\text{thick}}$ contains the same dashed lines as $\gamma$.
\item $\gamma_{\text{thick}}$ contains all vertices in $\gamma$ \emph{and} all vertices that are
connected to vertices in $\gamma$ by an edge of $\Lambda$.
\end{enumerate}
For later use we note that, since each vertex in $\Lambda$ is connected by edges to 14 other vertices,
we have the upper bound
\beq
	|\gamma_{\text{thick}}|\leq 14|\gamma|\ . 
\eeq

Using this definition of thickened polymers, we can bound a sum of the form \eqref{eq:overlap-sum}
for our model as
\beq
		\sum_{\substack{\gamma'\\\gamma' \cap \gamma \neq \varnothing}}|W(\gamma')|g(|\gamma'|) \leq \sum_{v \in\gamma_{\text{thick}}} \sum_{\gamma'\ni v}|W(\gamma')|g(|\gamma'|)\ ,
\eeq
where the outer sum is now taken over all $v$ in $\gamma_{\text{thick}}$. 
If we again assume a bound $|W(\gamma)| \leq \td{W}(\gamma)$, 
and that $\Gamma$ is isotropic with respect to $\td{W}(\gamma)$, then we can obtain a bound that looks similar to the one from the previous case
with the simpler notion of intersection,
\beqa
	\sum_{\substack{\gamma'\\\gamma' \cap \gamma \neq \varnothing}}|W(\gamma')|g(|\gamma'|) &\leq& \sum_{v \in\gamma_{\text{thick}}} \sum_{\gamma'\ni v}\td{W}(\gamma')g(|\gamma'|) \nnb \\
	&=& |\gamma_{\text{thick}}| \sum_{\gamma'\ni v}\td{W}(\gamma') g(|\gamma'|) \nnb \\
	&\leq& 14|\gamma|\sum_{\gamma'\ni v}\td{W}(\gamma') g(|\gamma'|)\ ,
\eeqa
where we used the inequality $|\gamma_{\text{thick}}|\leq 14|\gamma|$ in the last line.
If we now follow the proof in Ref.~\onlinecite{brydges}, and use this method to 
bound overlap sums like \eqref{eq:overlap-sum}, then we will recover the modified
convergence criterion that we used for our model in Sec.~\ref{sec:lemma2}.

\section{Setup and the proofs of Lemma 1 and Lemma 2 for the general model}
\label{app:generalizations}

In this appendix we summarize the changes that need to be made in our setup and
in the proofs of Lemma 1 and 2 in order to prove the more general result that we presented in Sec.~\ref{sec:generalizations}.

\subsection{Setup}

The main change to our setup is in the form of the support sets $X$. As before, these
still consist of boxes and plaquettes. However, since we are no longer restricted to the case of 
2-body interactions, the dashed lines that we previously included in the support sets are not
capable of capturing the structure of the interactions in our more general model. 
As we explain below, we replace
the dashed lines with a new kind of structure, namely \emph{subsets} of boxes or plaquettes. Thus, our
support sets will be built from boxes, plaquettes, and subsets.
Before explaining the form of these subsets, we first discuss the form of the boxes and
plaquettes for models in spatial dimensions larger than one.

For ease of visualization we consider the case of $D=2$ spatial dimensions. The spatial
lattice is a square lattice and the blocked spacetime lattice is a cubic\footnote{For simplicity,
in this discussion we ignore the different lattice spacing in the temporal direction.} lattice. The
boxes from the $D=1$ case now become cubes that have a vertical bond of the blocked spacetime lattice running 
up their center. The rule for assigning such a cube to the support set of a microscopic configuration
is the same as in the $D=1$ case: if a perturbation with support on site $\mb{r}$ acts within a certain time 
slice, then the box (now a cube) centered on $\mb{r}$ within that time slice is included in the support set.

Moving on to the plaquettes, in the $D=2$ case, the domain wall worldlines now live on 
\emph{vertical faces} of the blocked spacetime lattice. These vertical faces are the analogues, in 
$D=2$, of the plaquettes from the $D=1$ case. If a 
microscopic configuration has a domain wall worldline running all the way through a vertical face of the
blocked spacetime lattice, then we include that vertical face in the support set of that configuration. 

Next, we come to the description of the \emph{subsets}. Each
subset contains between $2$ and $K$ boxes in the same time slice, or between $2$ and $K$ plaquettes
in the same time slice. This rule comes from the fact that the interactions in our general
model act on a subset of up to $K$ lattice sites. In the two-body case a dashed
line connecting two boxes is equivalent to the subset containing the two boxes at its ends,
but for higher-body interactions we instead work with subsets that contain all
of the boxes that participate in an interaction term. While the dashed lines
were labeled by their length $r$, the possible subsets are labeled by $Y$ and a spacetime vector
$\mb{R}$, where $Y$ is one of the subsets with $|Y|\geq 2$ that contributes to $V_{\mb{0}}$, and 
$\mb{R}$ is the location of a vertical bond midpoint (for a subset of boxes) or
a plaquette center (for a subset of plaquettes). In what follows, it will be convenient to denote such
a subset by $Y_{\mb{R}}$ and to say that this subset has ``shape'' $Y$. 

One small technical point about these subsets is that, if we have a subset of plaquettes, then we 
require all of the plaquettes in this subset to be separated from each other by a 
Bravais lattice vector of the (hyper)cubic lattice. We require this because,
as in the $D=1$ case, $Z_{\pm}$ does not actually receive contributions from configurations consisting of 
subsets of plaquettes, and so we can choose the rule for assembling plaquettes into subsets to be
whatever is most convenient for our combinatorial analysis.

Finally, in this setting we define the connected support sets $\td{\chi}$ in exactly the same manner as 
before. We then define the weakly-connected support sets $\chi$ to be those support sets made up of several 
connected support sets $\td{\chi}_1,\td{\chi}_2,\dots$ together with subsets such that it is possible to get 
from one connected support set in $\chi$ to any other by hopping between sites belonging to the same subset. 
For any support set $X$ we define $s_Y(X)$ to be the number
of subsets of shape $Y$ in $X$, and then we define $s(X) = \sum_Y s_Y(X)$ to be the 
total number of subsets in $X$.

\subsection{Lemma 1}

We now discuss the changes to the proof of Lemma 1. In fact, this proof proceeds exactly as 
before, after we define some notation and explain how to use the new summability condition 
\eqref{eq:general-sum-bound}.

First, we perform all traces using a basis $|s,\pm\ran$ for $\mathcal{H}_{\pm}$
that is defined in a similar way as before (the tuple $s$ is again defined with a fixed value of one particular spin, say the spin at $\mb{r}=\mb{0}$).

Next, we expand each operator $V_Y$ as a sum over Pauli strings, 
\beq
	V_Y = \sum_{A}v^{(A)}_Y \sigma^A\ ,
\eeq
where the multi-index $A$ labels Pauli strings that act on the sites contained in $Y$, and the
$v^{(A)}_Y$ are coefficients. To be more
precise, $A$ is a $|Y|$-tuple of the form $A = (a_1,\dots,a_{|Y|})$ where each component 
$a_i$ takes values in the set $\{0,x,y,z\}$, and then $\sigma^A$ is defined to be the product
\beq
	\sigma^A := \sigma^{a_1}_{\mb{r}_1}\dots\sigma^{a_{|Y|}}_{\mb{r}_{|Y|}}\ ,
\eeq
where $\sigma^0 := \mathbb{I}$ is the $2\times 2$ identity matrix and $\mb{r}_1,\dots,\mb{r}_{|Y|}$
are the lattice sites contained in $Y$ (one of them is the origin $\mb{0}$). 
There are at most $4^K$ values of the multi-index $A$ since $|Y|\leq K$. 

For each $V_Y$ we now define an operator
$\text{abs}(V_Y)$ by replacing all $\sigma^z$'s in $V_Y$ by $\mathbb{I}$'s, replacing all 
$\sigma^y$'s in $V_Y$ by $\sigma^x$'s, and finally replacing the coefficients $v^{(A)}_Y$ by their
absolute values $|v^{(A)}_Y|$. The operator $\text{abs}(V_Y)$ has the important property that all of
its matrix elements in the basis $|s,+\ran$ of $\mathcal{H}_{+}$ are positive.

Given this definition of $\text{abs}(V_Y)$, we then define
numbers $f_Y$ and operators $\td{V}_Y$ by
\beq
	f_Y := ||\text{abs}(V_Y)|| \ \ ,\ \  \td{V}_Y := \frac{\text{abs}(V_Y)}{f_Y}\ ,
\eeq
where we assume that $f_Y\neq 0$. In what follows we also assume that
the $f_Y$ satisfy the bound $|f_Y|\leq 1$ for all $Y$, as the perturbation term can always be made larger
by increasing the value of $\lambda$.

If the $V_Y$ satisfy the summability condition \eqref{eq:general-sum-bound}, then the $f_Y$ satisfy 
the condition $\sum_Y f_Y \leq c'$, where $c' = 4^K c$, which more closely resembles the
original summability condition \eqref{eq:sum-bound} from our simple model. To prove this, note
first that $||\text{abs}(V_Y)||\leq \sum_A |v^{(A)}_Y|$, and that 
$|v^{(A)}_Y| \leq ||V_Y||$. (This last bound can be proven using the formula
$v^{(A)}_Y = \text{Tr}\{V_Y \sigma^A\}/2^{L^D}$, where the trace is taken over the full Hilbert space, which has dimension $2^{L^D}$). Then, since there are at most $4^K$ values of $A$, we find that
$||\text{abs}(V_Y)|| \leq 4^K ||V_Y||$ and the modified summability condition
$\sum_Y f_Y \leq 4^K c$ follows.

With this notation we can prove
Lemma 1 following almost the same steps as before. The main difference from
our previous proof in Appendix~\ref{app:weights} is that we must now consider the contributions
to $W_{\pm}(X)$ from the individual Pauli
strings $\sigma^A$ that contribute to each perturbation term $V_{\mb{r}}$, 
\beq
	V_{\mb{r}} = \sum_Y T_{\mb{r}}V_Y T_{\mb{r}}^{-1}= \sum_Y\sum_A v^{(A)}_Y T_{\mb{r}}\sigma^A T_{\mb{r}}^{-1}\ .
\eeq
In particular, when we construct a precise formula for $W_{\pm}(X)$ (i.e., the analogue of Eqs.~\eqref{eq:W-sum} 
and \eqref{eq:precise-weights}), we use an indicator function that is equal to one only if the translated Pauli 
strings $T_{\mb{r}}\sigma^A T_{\mb{r}}^{-1}$, which are labeled by triples $(\mb{r},Y,A)$,
are consistent with the support set $X$ (of course, this perturbation term should also be acting at a time 
that is consistent with $X$). 
This allows us to extract the exponential factors associated
with the plaquettes in $X$, and we can then bound the rest of the weight using the operators $\td{V}_Y$ in a 
similar way as before. We then find that the final bound on the weights takes the form
\beq
	|W_{\pm}(X)| \leq e^{-\mu |X|}\delta^{s(X)}\left[\prod_Y f_Y^{s_Y(X)} \right]\ , 
	\label{eq:general-weight-bound}
\eeq
where $|X| = b(X) + p(X)$ as before. Of course, the exact value of $\lambda_0$ for which 
Lemma 1 holds will change due to the new form of the interaction.

\subsection{Lemma 2}

We now discuss the changes in the proof of Lemma 2. The weights $\td{W}(\chi)$ that appear in this
lemma are now given by the right-hand side of \eqref{eq:general-weight-bound}, and we can
also map each support set $\chi$ to a polymer $\gamma$ in the same manner as before. 
In this case our strategy is again to build up each weakly-connected polymer $\gamma$ by adding subsets to a 
special basic set of weakly-connected polymers. In our original proof of Lemma 2 in
Appendix~\ref{app:combinatorics} these were the \emph{tree-like} polymers. In this more general case the 
correct objects are not actually trees, and so we refer to them as \emph{minimal} weakly-connected polymers. 
We say that a weakly-connected polymer is minimal if it becomes disconnected upon the removal of any subset 
that it contains.

The key to the proof is again a growing procedure that generates a minimal weakly-connected polymer 
from a set of more basic data. In this case the data is as follows. 
We again require an ordering of all the vertices in $\Lambda$. Next, we require these ingredients:
\begin{enumerate}
\item{An $N$-tuple of connected polymers $(\td{\gamma}_1,...,\td{\gamma}_N)$, each of which contain $v$. This collection of polymers should have total size $\ell$: i.e. $\sum_{i=1}^N\ell_i = \ell$ where $\ell_i = |\td{\gamma}_i|$.} 
\item{An $m$-tuple of vertices $(\ov{s}_1,\dots,\ov{s}_m)$, where $(N-1)/(K-1) \leq m \leq N-1$.}
\item{An $m$-tuple $(Y^{(1)},\dots,Y^{(m)})$ of subsets, where the subset $Y^{(k)}$ contains $n_k \leq K$ points.}
\item{The vertex $\ov{s}_k$ belongs to $\td{\gamma}_{i_k}$ for
some integer $i_k$ satisfying $1\leq i_k \leq 1 + \sum_{k'=1}^{k-1} (\check{n}_{k'}-1)$. (Here the integer $\check{n}_{k'}$ is defined below).
We again assume that the pairs $(\ov{s}_k,i_k)$ form an increasing 
sequence with respect to an ordering ``$\prec$'' defined by
$(\ov{s}_k,i_k) \prec (\ov{s}_{k'},i_{k'})$ if either (1) $i_k < i_{k'}$, or (2) $i_k=i_{k'}$ and 
$\ov{s}_{k'}$ has a larger number than 
$\ov{s}_k$ in the original ordering of the vertices of $\Lambda$.}
\end{enumerate}

Using these ingredients, the growing procedure now takes the
following form. First, let $v_1 = v$ and $s_1 = \ov{s}_1$. Also, let 
$Y^{(1)}_{s_1}$ be the
translation of the subset $Y^{(1)}$ to the vertex $s_1$, and let 
$\{y^{(1)}_1,\dots,y^{(1)}_{n_1-1}\}$ be the $n_1-1$ vertices in 
$Y^{(1)}_{s_1}$ that are \emph{not} equal to $s_1$ (which came from the point at the origin $(\mb{0},0)$ 
of spacetime in the untranslated subset $Y^{(1)}$). We have also ordered these vertices
according to our original ordering of the vertices of $\Lambda$.
We now define $\check{n}_1 - 1$ vertices $v_2,\dots,v_{\check{n}_1}$, where $\check{n}_1 \leq n_1$, to be
equal to the subset of the vertices $\{y^{(1)}_1,\dots,y^{(1)}_{n_1-1}\}$
that are not contained in the connected polymer $\td{\gamma}_1$ (we also preserve the ordering when
defining these vertices). We then translate $\td{\gamma}_2$ by $v_2 - v$, 
$\td{\gamma}_3$ by $v_3 - v$, and so on, ending with a translation of 
$\td{\gamma}_{\check{n}_1}$ by $v_{\check{n}_1} - v$.
At this point the minimal weakly-connected polymer that we are
building consists of the union of translated polymers
\begin{align}
 \bigcup_{k=1}^{\check{n}_1} [\td{\gamma}_k + v_k - v] \label{eq:polymers-after-step-1}
\end{align}
together with the subset $Y^{(1)}_{s_1}$.

Next, we define a new vertex $s_2$ by the relation $s_2 = \ov{s}_2 + (v_{i_2} - v)$. In addition,
let $Y^{(2)}_{s_2}$ be the translation of $Y^{(2)}$ to the vertex $s_2$, and let 
$\{y^{(2)}_1,\dots,y^{(2)}_{n_2-1}\}$ be the $n_2-1$ vertices in 
$Y^{(2)}_{s_2}$ that are not equal to $s_2$ (and we again order these according to the original ordering on 
$\Lambda$). We then define $\check{n}_2-1$ new vertices $v_{\check{n}_1+1},v_{\check{n}_1+2},\dots,
v_{\check{n}_1+\td{n}_2-1}$,
where $\check{n}_2\leq n_2$, to be equal to the subset of the vertices 
$\{y^{(2)}_1,\dots,y^{(2)}_{n_2-1}\}$
that are not contained in the union \eqref{eq:polymers-after-step-1} that we have
already constructed. We then add the translated polymers
$\bigcup_{k=2}^{\check{n}_2} [\td{\gamma}_{\check{n}_1 + k - 1} + v_{\check{n}_1 + k  -1} - v]$ and
the subset $Y^{(2)}_{s_2}$ 
to the minimal weakly-connected polymer that we are growing. At this point our
minimal weakly-connected polymer consists of the union 
\begin{align}
 \bigcup_{k=1}^{\check{n}_1 + \check{n}_2 - 1} [\td{\gamma}_k + v_k - v] \nnb
\end{align}
together with the subsets $Y^{(1)}_{s_1}$ and $Y^{(2)}_{s_2}$. 

We now continue the construction in this manner until all of the original connected polymers
$\td{\gamma}_1,\dots,\td{\gamma}_N$ are used up. As in our original growing procedure
from Appendix~\ref{app:combinatorics}, this procedure is capable of growing all minimal weakly-connected
polymers (and many others as well), and so it leads to an upper bound on the number of minimal
weakly-connected polymers that contain the distinguished vertex $v$. 
Let $q_{\text{min}}(N,\ell)$ be the following sum over all minimal $\gamma$ with $N$ connected components 
and total size $\ell$:
\begin{align}
	q_{\text{min}}(N,\ell) =  \sum_{\substack{\gamma\ni v\\ \text{$\gamma$ minimal}\\ n_c(\gamma)=N \\ |\gamma|=\ell}}\td{W}(\gamma)e^{14|\gamma|}\ .
\end{align}
Then our growing procedure leads to an upper bound on this quantity of the form
\beq
	q_{\text{min}}(N,\ell)\leq \sum_{\substack{\ell_1,\dots,\ell_N \\ \ell_1+\dots+\ell_N=\ell}}\prod_{i=1}^N q_0(\ell_i) \sum_{m = (N-1)/(K-1)}^{N-1}\sum_{\substack{\ov{s}_1,\dots,\ov{s}_{m} \\ (\ov{s}_k,i_k) \text{increasing} \\
	\text{w.r.t. ``$\prec$''}}}\sum_{Y^{(1)},\dots,Y^{(m)}}\delta^m\prod_{k=1}^{m}f_{Y^{(k)}}\ ,
\eeq
where the $q_0(\ell_i)$ are again the contributions from the $\td{\gamma}_i$. The main difference
between this expression and Eq.~\eqref{eq:growing-procedure-bound} is that we now have an extra sum over $m$, 
which counts the number of subsets that are needed to build up each minimal configuration (in the case of 
two-body interactions, where $K=2$, we must have $m=N-1$ to build a tree-like polymer out of $N$ connected 
components). As before, we also find that $q(N,\ell)\leq e^{\ell\delta c'}q_{\text{min}}(N,\ell)$
(recall that $c' = 4^K c$)  by
the same argument of adding or not adding every possible subset (and then using the summability condition
for the $f_Y$).

Finally, we proceed to bound $q = \sum_{\ell=1}^{\infty}\sum_{N=1}^{\ell}q(N,\ell)$ as before. The main
new feature that we encounter here is the sum
\beq
	\sum_{N=1}^{\ell}\sum_{m = (N-1)/(K-1)}^{N-1}\binom{2\ell}{m}(\delta c')^m\ , \nnb
\eeq
which can be bounded as follows, 
\begin{align}
	\sum_{N=1}^{\ell}\sum_{m = (N-1)/(K-1)}^{N-1}\binom{2\ell}{m}(\delta c')^m &\leq \sum_{N=1}^{\ell}\sum_{m=0}^{2\ell}\binom{2\ell}{m}(\delta c')^m \nnb \\
	 &= \ell(1+\delta c')^{2\ell} \nnb \\
	&\leq e^{(2\delta c' + 1)\ell}\ .\nnb 
\end{align}
We then complete the proof in the same way as before.


%

\end{document}